%% file: paper.tex
\documentclass[sigconf]{acmart}

\usepackage[frozencache,cachedir=_minted-paper]{minted}
\usepackage{xspace}
\usepackage{booktabs}
\usepackage{multirow}
\usepackage{graphicx}
\usepackage{subcaption}
\usepackage{microtype}
\usepackage{tikz}
\usepackage{tcolorbox}
\tcbuselibrary{breakable}
\usepackage{colortbl}
\usepackage{varwidth}
\usepackage{framed}
\usepackage{tabularx}

\input{macros}

\newcommand{\tool}{\textsc{Gentoo}\xspace}
\newcommand{\toolStatic}{\textsc{Gentoo-S}\xspace}
\newcommand{\toolLLM}{\textsc{Gentoo-L}\xspace}
\newcommand{\toolBase}{\textsc{Gentoo-Base}\xspace}

\title{Fuzzing with Agents? Generators Are All You Need}

\author{Vasudev Vikram}
\affiliation{%
  \institution{Carnegie Mellon University}
  \city{Pittsburgh}
  \state{PA}
  \country{USA}
}
\email{vasumv@cmu.edu}

\author{Rohan Padhye}
\affiliation{%
  \institution{Carnegie Mellon University}
  \city{Pittsburgh}
  \state{PA}
  \country{USA}
}
\email{rohanpadhye@cmu.edu}

\begin{abstract}
Modern generator-based fuzzing techniques combine lightweight input generators with coverage-guided mutation as a method of exploring deep execution paths in a target program. A complimentary approach in prior research focuses on creating highly customized, domain-specific generators that encode structural and semantic logic sufficient enough to reach deep program states; the challenge comes from the overhead of writing and testing these complex generators. We investigate whether AI coding agents can automatically synthesize such
target-specific generators, and whether the resulting generators are strong
enough to obviate the need for coverage guidance and mutation entirely.

Our approach, \tool, is comprised of an LLM coding
agent (provided terminal access and source code of the fuzz target and its library) instructed to iteratively
synthesize and refine an input generator, and optionally provided fine-grained predicate-level
coverage feedback.
We evaluate three configurations of \tool against human-written generators on
fuzz targets for 7 real-world Java libraries.
Our findings show that agent-synthesized generators achieve statistically
significantly higher branch coverage than human-written baseline generators on 4 of 7
benchmarks.
Critically, the use of coverage guidance and mutation strategies is \textit{not} 
statistically significantly beneficial for agent-synthesized generators, but is
significant for all human-written generators, suggesting that structural and
semantic logic encoded in the agent generators makes coverage guidance largely
unnecessary.
\end{abstract}

\begin{document}

\maketitle

\input{intro}
\input{background}

\input{tech}
\input{eval}
\input{related}
\input{threats}
\input{discussion}

\bibliographystyle{ACM-Reference-Format}
\bibliography{refs}

\end{document}

%% file: macros.tex
\newcommand{\code}{\texttt}

 \definecolor{vvcolor}{rgb}{0.1,0.5,0.1}    
\newcommand{\DIFdel}[1]{} 

\newcommand*{\mybox}[1]{\framebox{#1}}
\definecolor{c3}{HTML}{eaeccc}
\definecolor{c2}{HTML}{33bbee}
\definecolor{c1}{HTML}{44AA99}
\definecolor{c4}{HTML}{98cae1}
\definecolor{c5}{HTML}{cc3311}
\definecolor{c6}{HTML}{FFFFFF}
\definecolor{c7}{HTML}{FF0000}

\newcommand{\rom}[1]{\uppercase\expandafter{\romannumeral #1\relax}}

\usepackage{varwidth}

\colorlet{grey}{red!40}

\def\mystrut(#1,#2,#3){\vrule height #1pt depth #2pt width #3pt}    

\definecolor{boxcolor}{RGB}{238, 223, 204} %
\DeclareRobustCommand{\mybox}[2][gray!20]{%
\begin{tcolorbox}[   
        breakable,
        left=0pt,
        right=0pt,
        top=0pt,
        bottom=0pt,
        colback=#1,
        colframe=black,
        width=\dimexpr\columnwidth\relax, 
        enlarge left by=0mm,
        boxsep=5pt,
        outer arc=4pt,
        boxrule=.5mm
        ]
        #2
\end{tcolorbox}
}

\definecolor{lightyellow}{RGB}{255,255,191}
\definecolor{customred}{RGB}{255,0,0}

%% file: intro.tex
\section{Introduction}
\label{sec:intro}

Fuzz testing is a widely used technique for finding bugs and vulnerabilities in software by
repeatedly executing a program with automatically generated random inputs~\cite{Miller90,Rebert14,Cha15,Chen18-angora,Aschermann19-redqueen,Lyu19,Gan20,Ding21,Klees18}.
A fundamental challenge in fuzz testing is generating inputs that are both structurally valid and semantically meaningful: inputs that are too random are rejected early by validation checks, while inputs that are too simple fail to reach the deep program logic where interesting bugs live.
This challenge motivated greybox fuzzing~\cite{Afl}, in which a corpus of inputs is gradually mutated and evolved, saving new input candidates that
increase overall branch coverage and discarding those that do not.
By making small, incremental changes to known "interesting" inputs, greybox fuzzing can
efficiently reach deep program states well beyond validation
checks~\cite{Bohme16,Valentin19,Lemieux18,Bohme22,Padhye19-jqf,Lampropoulos19}.

\begin{figure}[t]
  \centering
  \includegraphics[width=\linewidth]{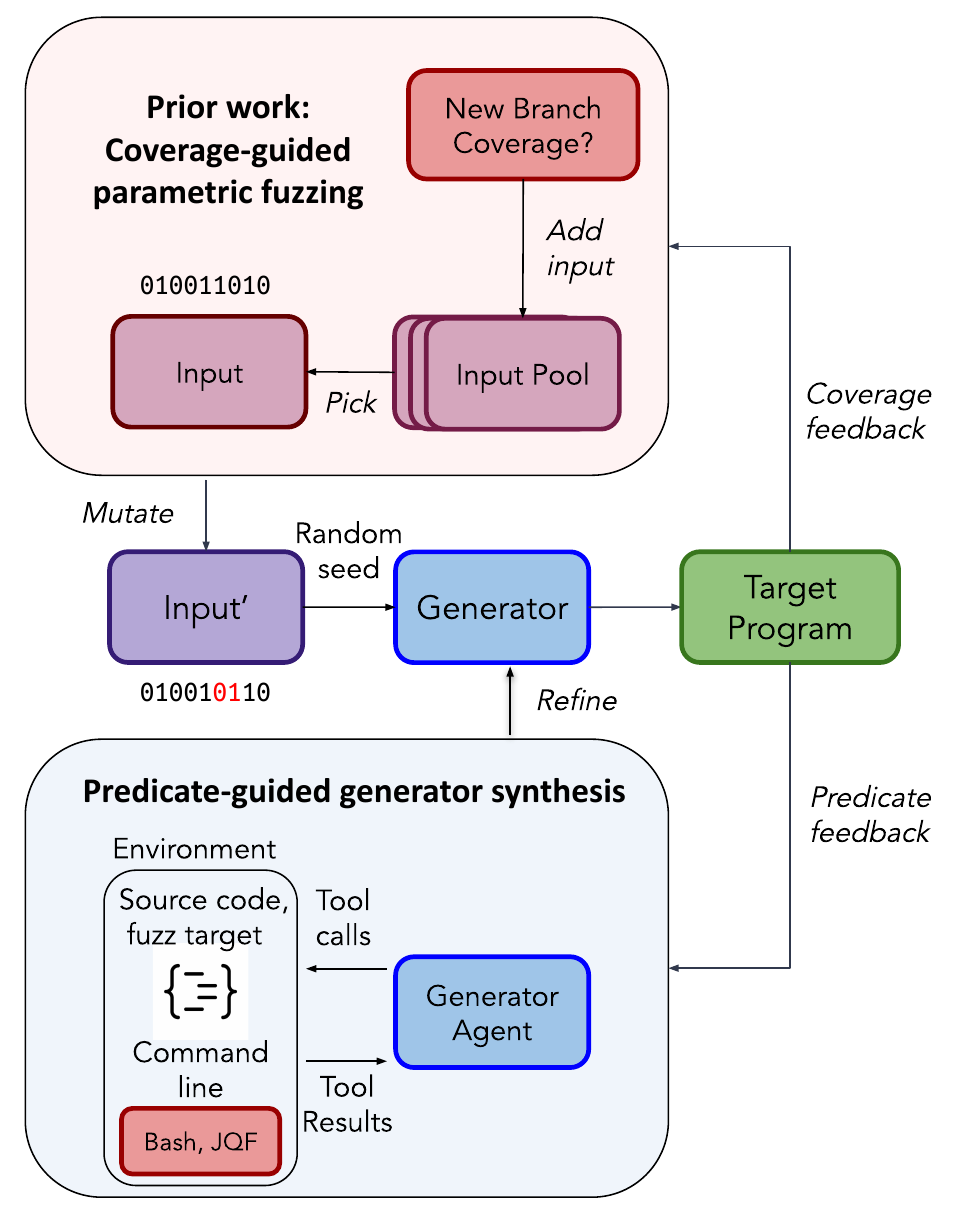}
  \caption{Comparison of prior approaches (coverage-guided parametric fuzzing) with our proposed approach \tool: agentic synthesis of input generators with fine-grained predicate feedback.}
  \label{fig:gentoo-overview}
\end{figure}

Parametric generator-based fuzzing~\cite{Padhye19-zest, Padhye19-jqf, zeugma, nguyen2022bedivfuzz} combines coverage guidance and mutation-based fuzzing with lightweight generator functions that produce structurally valid inputs for a given target program. JQF~\cite{Padhye19-jqf} and Zest~\cite{Padhye19-zest} showed this is possible by treating generators as \emph{parametric decoders}: functions that consume an arbitrary byte sequence to produce a structured input, so that byte-level mutations from a coverage-guided fuzzer translate directly into structured mutations on the generated input without requiring specialized mutators.

A complementary line of work focuses on building heavyweight, highly customized, domain-specific generators that encode hard input constraints directly into the generation logic.
Tools such as CSmith~\cite{Yang11, Regehr12} for C compilers and SQLancer~\cite{ba2023testing} for database engines generate inputs that are not merely syntactically valid but
semantically rich by construction: CSmith emits C programs free of undefined
behavior, while SQLancer produces queries that satisfy the relational schemas and
type rules of the target database. By encoding deep input and domain logic into the generators, these tools routinely expose bugs that generic mutation-based fuzzers miss entirely, because the bugs only manifest on inputs that satisfy constraints random mutations are less likely to produce.
Notably, such heavyweight generators do \emph{not} employ coverage guidance nor mutation strategies, since the generation logic itself is strong enough that neither is necessary.
The limitation, however, is that such generators require substantial manual effort to write and are tightly coupled to a single target.

With the rise of large language models (LLMs) capable of automating complex software engineering tasks~\cite{xia2025livesweagentsoftwareengineeringagents, yang2024swe, jimenez2023swe, liu2025largelanguagemodelbasedagents, guo2025comprehensivesurveybenchmarkssolutions, vikram2023can, maaz2025agentic}, a natural instinct is to apply them to
fuzz testing and input generation tasks.
Recent work has used LLMs to synthesize fuzz drivers~\cite{lyu2023prompt, xu2024ckgfuzzer, liu2025promefuzz, ossfuzz23} and input generators~\cite{chen2025elfuzz, zhang2025low}, demonstrating that LLMs can produce generation logic that improves
fuzz testing applicability and performance.
However, these approaches use LLMs to improve components within the mutation-based fuzzing paradigm, still operating on the assumption that the coverage guidance loop itself is necessary.
We believe synthesizing input generators is a well-suited task for AI coding \textit{agents}, requiring a deep understanding of source code and the ability to translate program semantics into input generation logic.
In particular, we are focused on the following motivating question: \emph{can agentic synthesis of customized input generators obviate the need for coverage guidance and mutation-based fuzzing?}

\subsection{Motivating Example}

Consider the type-checking logic in the ChocoPy compiler~\cite{Padhye19-chocopy}, which compiles programs from a typed subset of Python into RISC-V assembly. To exercise the type-checking
logic beyond its entry point, the fuzzer must produce inputs that are not only
syntactically valid programs but also well-typed ones, a constraint that no generic
mutation strategy can efficiently satisfy without understanding the target's semantics.

Figure~\ref{fig:chocopy-example} shows a simple ChocoPy program. ChocoPy is a
statically typed subset of Python where every variable and function signature
carries an explicit type annotation.
Figure~\ref{fig:chocopy-predicate} shows two specific branches from the ChocoPy compiler
\texttt{DeclarationAnalyzer} that must be reached to exercise deep type-checking
logic.
The predicate on line~3 (\texttt{parentMethodDecl != null}) is reached only when
the input program defines a class that \emph{inherits} from another class
\emph{and} overrides one of its methods: the fuzzer must generate at least two
classes in an inheritance relationship with a method of the same name in both.
Even if a grammar-based sampler produces a syntactically valid ChocoPy program,
the probability that it happens to include an inheritance hierarchy with a method
override is very low.
The predicate on line~5 (\texttt{!signatureMatches}) is reached only when that
overriding method has a \emph{different} type signature from the parent's
method.
We hypothesize that a specialized input generator should encode this logic
directly, generating ChocoPy programs with inheritance hierarchies and method
overrides by construction, rather than relying on coverage guidance and
mutations to produce them.


\begin{figure}[t]
  \begin{subfigure}[t]{\linewidth}
\begin{minted}[fontsize=\footnotesize,frame=single,breaklines]{python}
# Search in a list
def contains(items:[int], x:int) -> bool:
    i:int = 0
    while i < len(items):
        if items[i] == x:
            return True
        i = i + 1
    return False
\end{minted}
    \caption{A simple ChocoPy program. ChocoPy is a statically typed subset of Python
      where all variables and function signatures carry explicit type annotations.}
    \label{fig:chocopy-example}
  \end{subfigure}

  \medskip

  \begin{subfigure}[t]{\linewidth}
\begin{minted}[fontsize=\footnotesize,frame=single,breaklines,linenos]{java}
// Reached only if input declares a class that inherits
// from another class AND redefines one of its methods
if (parentMethodDecl != null) {          // predicate 1
    // ... check parameter types and return type match ...
    if (!signatureMatches) {             // predicate 2
        err(methodName,
            "Method overridden with different type signature: %s",
            methodName.name);
    }
}
\end{minted}
    \caption{Two example branches from the ChocoPy compiler \texttt{DeclarationAnalyzer} Java source code (simplified).
      Line~3 requires an input with an inheritance hierarchy and a method override;
      line~5 additionally requires the override to have a mismatched type signature.}
    \label{fig:chocopy-predicate}
  \end{subfigure}
  \label{fig:chocopy}
\end{figure}

\subsection{Contributions}

We investigate the ability of agentic methods to synthesize target-specific input
generators for fuzz testing.
Effective synthesis requires: (1) giving the agent the tooling and source context
to write a generator, (2) measuring the gap between what the current generator
produces and what the target's deep predicates require, and (3) iterating and
refining the generator to close that gap.

The mechanism we choose to measure and close this gap is \emph{predicate feedback}.
On the static side, we apply interprocedural control-flow analysis to rank all
conditional branches in the library under test by their branch dominance score,
identifying the predicates whose outcomes unlock the most unexplored code.
On the dynamic side, we instrument the fuzzer to track, for each saved input,
which branch each tracked predicate takes, producing a per-predicate coverage
report that tells the agent not just whether a predicate was reached but how
consistently its harder branch is satisfied.
We name the resulting system \tool (\textbf{gen}erate the \textbf{gen}erator,
shown in Figure~\ref{fig:gentoo-overview}), and evaluate three configurations:
\toolStatic, which uses static dominator analysis to rank predicates;
\toolLLM, which instead asks the agent to identify and rank predicates from
source code; and \toolBase, a plain coding agent with no predicate guidance.

We evaluate all three configurations on seven real-world Java benchmarks spanning
bytecode parsing, ASN.1 cryptographic structures, JavaScript compilation and
interpretation, BZip2 compression, JSON parsing, and ChocoPy compilation.
For each benchmark we compare against human-written generators from prior
work~\cite{Padhye19-zest, mu2, Vikram21} and measure both coverage-guided and
random fuzzing to assess whether the synthesized generators reduce the need for
mutation-based exploration.

Our evaluation yields three main takeaways.
First, agent-synthesized generators achieve higher branch coverage than
human-written generators on 6 of 7 benchmarks (4 statistically significant),
by $11$--$21\%$ on average,
suggesting that agentic synthesis is a viable and competitive alternative
to manual generator development.
Second, running the synthesized generators under coverage-guided fuzzing
yields marginal improvements over random sampling (typically under 3\%),
and these differences are not statistically significant, supporting
the hypothesis that a well-synthesized generator can largely stand on its own
without coverage-guided mutation.
Third, predicate feedback measurably refines generators across iterations --
\toolStatic and \toolLLM improve over their first-iteration baselines by
$18$--$57\%$ on average respectively -- yet final coverage is broadly
comparable across all three \tool configurations, indicating that the
initial agent synthesis accounts for most of the coverage gain and
predicate feedback provides an additional but not dominant boost.

%% file: background.tex
\section{Background}
\label{sec:background}

\subsection{Coverage-Guided Fuzzing}

Coverage-guided greybox fuzzing (CGF) is a technique for automated test-input
generation using lightweight program instrumentation, popularized by
AFL~\cite{Afl}. As shown in the top half of Figure~\ref{fig:gentoo-overview}, the fuzzer maintains an \emph{input pool} (or corpus) of saved inputs.
At each iteration, a seed input is selected from the pool, typically according
to an energy schedule that favors seeds that have historically produced
coverage-increasing mutations~\cite{aflfast}.
The selected seed is then \emph{mutated}, commonly by applying random byte-level
operations such as bit flips, byte substitutions, or block insertions and
deletions, to produce a new candidate input $x'$.
The program is executed on $x'$ under instrumentation that records the set of
branches exercised during the run.
If $x'$ exercises at least one branch not covered by any input already in the
pool, it is added to the pool and becomes eligible for future mutation.
This saving criterion ensures the pool grows incrementally toward broader
coverage, while inputs that do not increase coverage are discarded.

The effectiveness of this feedback loop depends critically on the ability of
byte-level mutations to produce inputs that satisfy the program's structural
constraints. For programs that accept highly structured inputs, such as compilers, parsers,
or serialization libraries, most random byte mutations produce malformed inputs
that fail early validation, preventing the fuzzer from reaching the deeper
program logic where interesting bugs tend to live.

\subsection{Parametric Generator-Based Fuzzing}

Property-based testing frameworks such as QuickCheck~\cite{claessen2000quickcheck, junit-quickcheck, junit-quickcheck-implied} ask the
developer to supply a \emph{generator}: a function that produces random,
structurally valid values by making a sequence of random choices, such as
selecting an element from a set or deciding the length of a list.
Generators can encode arbitrarily rich structural invariants, ensuring that every
generated input satisfies the constraints required to pass shallow validation
and reach deeper program logic.
For example, a JQF generator for ChocoPy programs might look like
Figure~\ref{fig:jqf-generator}.

\begin{figure}[t]
\begin{minted}[fontsize=\footnotesize,frame=single,breaklines]{java}
String generate(SourceOfRandomness random, ...) {
    String program = "";
    // int numFunctions = random.nextInt(1, 5);
    // for each function: generate signature, typed params, body...
    // program += generateFunction(random);
    // ...
    return program;
}
\end{minted}
\caption{Skeleton of a JQF \texttt{Generator} for ChocoPy. The generator
  uses \texttt{SourceOfRandomness} to make random choices that produce
  a structured string input.}
\label{fig:jqf-generator}
\end{figure}

Zest~\cite{Padhye19-zest} and JQF~\cite{Padhye19-jqf} bridge generator-based and coverage-guided fuzzing through the
concept of a \emph{parametric generator}.
The key idea is to back the random source used by a generator with an explicit
byte stream $B$.
Each call to the random API (e.g., \texttt{nextBoolean()}, \texttt{nextInt()})
consumes one or more bytes from $B$ and uses their values to resolve the
corresponding generation decision.
The generator is then deterministic given $B$: replaying the same stream always
produces the same structured input.
This makes the byte stream an underlying representation of the generated
input, and byte-level mutations on $B$ correspond directly to structured
mutations on the generated output without requiring specialized mutators.
A coverage-guided fuzzer can therefore operate on $B$ as it would on any
raw input, applying its existing mutation operators and saving criterion,
while the generator deterministically decodes $B$ into a structurally valid input
at each execution.

Despite this elegant combination, parametric generators are subject to the
\emph{havoc effect}~\cite{havoc}: mutations to bytes that influence early
branching decisions in the generator can drastically restructure the generated
input, causing large jumps in the structured input space that undermine the
incremental exploration on which coverage guidance relies.
Even so, the lightweight generators combined with coverage guidance represent a
substantial improvement over raw byte mutation for structured inputs~\cite{Padhye19-zest}.
Understanding which program predicates block deeper exploration, and targeting
generation toward satisfying them, is the natural next step.

\subsection{Control-Flow Analysis and Dominators}

To identify which predicates are worth adding custom generation logic for,
we need a principled way to rank them by how much unexplored code they
control.

A program's \emph{control-flow graph} (CFG) is a directed graph where nodes
represent basic blocks and edges represent possible transfers of control.
A \emph{predicate} is a conditional branch with two (or more) successor blocks,
each corresponding to a possible outcome of the branch condition.
Predicates act as gatekeepers: a predicate that must be satisfied to reach
some region of the program blocks all coverage of that region until an input
satisfying it is found.

The gatekeeping relationships in a CFG are formalized by the
\emph{dominator tree}.
A node $d$ \emph{dominates} node $n$ if every path from the entry of the CFG to
$n$ passes through $d$.
The dominator tree organizes these relationships hierarchically: $d$ is the
parent of $n$ in the tree if $d$ is the \emph{immediate dominator} of $n$,
the closest strict dominator of $n$ along any entry-to-$n$ path.

We define the \emph{branch dominance score} to quantify how much code each
outcome of a predicate controls.
Let $p$ be a predicate with successor blocks $b_1, b_2, \ldots$\,.
For each successor $b_i$, let $R(b_i)$ be the set of CFG nodes reachable from
$b_i$ and let $D(b_i)$ be the set of CFG nodes dominated by $b_i$ in the
dominator tree.
The branch dominance score of outcome $b_i$ is
\[
  \text{dom}(b_i) = |R(b_i) \cap D(b_i)|,
\]
the number of nodes that are both exclusively controlled by taking branch $b_i$
and reachable from it.
A high branch dominance score indicates that a particular outcome of the
predicate unlocks a large region of otherwise unreachable program logic.
The overall dominance score of the predicate $p$ is
$\text{dom}(p) = \max_i \text{dom}(b_i)$.

By ranking predicates by dominance score, we can identify the branch conditions
most worth targeting: those whose unsatisfied outcomes block the broadest
regions of the program.

%% file: tech.tex
\section{Approach}
\label{sec:tech}

We investigate the ability of agentic methods to synthesize target-specific
input generators for fuzz testing.
The core idea is to give a coding agent three things: the source context and
tooling needed to write a generator, a way to measure the gap between what the
generator currently produces and what the target requires, and the ability to
iterate and refine the generator to close that gap.
Our system \tool (\textbf{gen}erate the \textbf{gen}erator), shown in
Figure~\ref{fig:gentoo-overview}, produces a JQF input generator via an
iterative agent loop.

We evaluate three configurations of \tool that differ in how the gap is
measured and communicated to the agent.
\toolBase provides the agent with terminal access to inspect JQF execution
statistics and examine generated inputs, without any additional analysis; the
agent infers what to improve from the raw fuzzing output alone.
\toolStatic and \toolLLM both go further by providing predicate-level feedback
that directly measures which branches of high-value predicates the current
generator fails to reach. They differ in how the important predicates are
identified: \toolStatic uses static interprocedural dominator analysis to rank
predicates by the amount of code each branch controls, while \toolLLM instead
asks the agent to analyze the library source code and identify the predicates
it judges most important.


\subsection{Dominator Predicate Analysis}
\label{sec:analysis}

The first strategy for identifying high-value predicates, used by \toolStatic,
is static interprocedural CFG analysis.
The fuzz target defines one or more entry points into the library under
test -- the library methods that the target invokes on each generated input.
Starting from these entry points, \tool constructs an interprocedural
CFG\footnote{Each benchmark has an exclusions list to filter out non-application-specific
classes, such as standard \texttt{java.lang} classes, keeping the analysis focused
on library code.} and computes its dominator tree.
For each conditional branch in this CFG, \tool computes the branch dominance
score $\text{dom}(b_i) = |R(b_i) \cap D(b_i)|$ from Section~\ref{sec:background},
where $R(b_i)$ is the set of CFG nodes reachable from branch outcome $b_i$ and
$D(b_i)$ is the set of nodes dominated by $b_i$.

The output is a ranked list of predicate records, sorted by the maximum
branch dominance score across all outcomes of each predicate.
Each predicate record identifies the predicate by its source class, method, and
line number, and lists each of its branch outcomes with its individual dominance
score.
This list is passed to the agent as the primary guide for what generation logic
to add: predicates at the top of the ranking gate the most code, so satisfying
their harder-to-reach branch outcomes unlocks the most additional coverage.
Figure~\ref{fig:predicates-and-coverage}a shows an example predicate record
from ChocoPy's \texttt{TypeChecker}, with overall dominance score 50 and two
branches with scores 10 and 40 respectively.
We implement this analysis using the WALA static analysis framework~\cite{wala}.

\subsubsection{LLM Predicate Identification}
\label{sec:llm-analysis}

As an alternative used by \toolLLM, the agent itself scans the library
source code and produces a ranked predicate list without running a static
analyzer.
Given the fuzz target and library source, the agent reads relevant classes,
identifies conditional statements whose branches it judges to require non-trivial
inputs to satisfy, and orders them by estimated importance.
This approach requires no pre-built static analysis tooling and leverages the
agent's natural-language understanding of the code's intent.

Figure~\ref{fig:predicates-and-coverage} illustrates the two data structures
passed to the agent for a predicate in ChocoPy's \texttt{TypeChecker}.
The predicate at line~206 of \texttt{analyzePlus} is \texttt{leftType.equals(rightType)},
the type check for the \texttt{+} operator.
Its branch at line~207 (taken when both operands share the same type, e.g.\
\texttt{int + int}) controls only 10 nodes, while the branch at line~208
(taken when both operands are lists of different element types,
triggering least-upper-bound computation) controls 40 nodes.
After fuzzing, the dynamic record reveals that 798 inputs took the low-value
branch (line~207) while only 24 took the high-value branch (line~208), out of
822 total inputs that reached this predicate.
To reach line~208 the generator must produce a ChocoPy program that applies
\texttt{+} to two lists with different element types (e.g.\ \texttt{[1,2] + ["hello"]}),
a specific semantic constraint that a generic generator almost never satisfies by chance.

\begin{figure}[t]
\begin{subfigure}[t]{\linewidth}
\begin{minted}[fontsize=\footnotesize,frame=single,breaklines]{json}
{
  "class": "...TypeChecker",
  "method": "analyzePlus",
  "line": 206,
  "branches": [
    { "line": 207, "dominance": 10 },
    { "line": 208, "dominance": 40 }
  ]
}
\end{minted}
\caption{Static predicate record: source location,
  overall dominance score, and per-branch dominance scores.}
\end{subfigure}

\medskip

\begin{subfigure}[t]{\linewidth}
\begin{minted}[fontsize=\footnotesize,frame=single,breaklines]{json}
{
  "class": "...TypeChecker",
  "method": "analyzePlus",
  "predicateLine": 206,
  "predicateInputs": 822,
  "branches": [
    { "line": 207, "inputs": 798 },
    { "line": 208, "inputs": 24  }
  ]
}
\end{minted}
\caption{Dynamic predicate record: per-branch
  input counts after a fuzzing campaign. The high-dominance branch at line~208
  was taken by only 24 of 822 inputs that reached this predicate.}
\end{subfigure}

\caption{Static (a) and dynamic (b) predicate records for a predicate in
  ChocoPy's \texttt{TypeChecker}. The static record identifies that line~208
  is the high-value branch; the dynamic record reveals the generator rarely
  produces inputs that take it, making it a direct target for refinement.}
\label{fig:predicates-and-coverage}
\end{figure}

\subsection{Dynamic Predicate Feedback}
\label{sec:dynamic}

Both \toolStatic and \toolLLM provide the agent with fine-grained per-predicate
feedback by adding instrumentation on top of the standard coverage-guided fuzzer.
At each conditional branch corresponding to a tracked predicate, the fuzzer
records which branch outcome each saved input takes.
After the campaign, this produces a \emph{predicate coverage report} that
summarizes, for each predicate and each branch outcome, the number of distinct
saved inputs that triggered that branch.

This per-branch view goes beyond aggregate branch coverage, which only tells
the agent \emph{whether} a predicate was ever reached.
The predicate coverage report additionally reveals \emph{how consistently}
the generator produces inputs that satisfy each branch condition.
A predicate with a high dominance score whose harder branch is hit by very few
inputs is a direct target for refinement: it gates a large amount of code, and
the current generator rarely produces inputs that unlock it.
Figure~\ref{fig:predicates-and-coverage}b shows the dynamic record for the
same \texttt{TypeChecker} predicate: 798 inputs took the low-value branch
(line~207) and only 24 took the high-value branch (line~208) that controls
40 nodes, making it an obvious refinement target.

\subsection{Agent-based Generator Synthesis}
\label{sec:synthesis}
\begin{table*}[t]
\small
\centering
\caption{Comparison of \tool configurations. All share terminal access,
  file editing, library source code, and JQF tooling (compile, fuzz, corpus replay).}
\label{fig:agent-instructions}
\begin{tabularx}{\linewidth}{l l X}
\toprule
\textbf{Config} & \textbf{Predicate identification} & \textbf{System prompt (summary)} \\
\midrule
\toolBase
  & None
  & Read source, write generator, refine based on JQF stats and corpus inspection. \\
\addlinespace
\toolLLM
  & LLM source analysis
  & $+$ Identify important predicates from source and produce a predicate record; use the predicate coverage report after each run to identify uncovered high-dominance branches and refine the generator. \\
\addlinespace
\toolStatic
  & Static dominator analysis
  & $+$ Read the pre-computed predicate record; use the predicate coverage report after each run to identify uncovered high-dominance branches and refine the generator. \\
\bottomrule
\end{tabularx}
\end{table*}

All three configurations drive generator synthesis using the Claude Agent
SDK~\cite{claude-agent-sdk}, which provides the agent with terminal access and
file editing capabilities (Table~\ref{fig:agent-instructions}).
All configurations receive the same base instructions: read through the fuzz
target and library source code, synthesize a JQF-style \texttt{Generator} class
targeting maximum branch coverage, and iteratively refine it.
To support the refinement loop, we provide the agent with additional
documentation on using JQF via the terminal to run short-scale fuzzing sessions
(2 minutes) and to replay the saved corpus of inputs, allowing the agent to
directly inspect the inputs its current generator produces.

Each configuration additionally receives feedback tailored to its strategy.
For \toolLLM, the agent is instructed to scan the library source code, identify
the conditional predicates it judges most important for coverage, and produce
a predicate records file in the format described in Section~\ref{sec:llm-analysis}.
For \toolStatic, the agent is pointed to the pre-computed predicate records file
produced by the dominator analysis (Section~\ref{sec:analysis}).
In both cases, the agent uses the predicate records alongside the dynamic
per-branch coverage report from the previous fuzzing session to identify which
high-value predicates remain underserved and revise the generator accordingly.
\toolBase receives neither; it refines the generator based solely on JQF
execution statistics and its own inspection of the generated inputs.

We deliberately chose this lightweight, freeform approach rather than enforcing
a rigid synthesis pipeline.
By giving the agent terminal access and open-ended instructions, it can
interleave reading source code, inspecting generated inputs, running fuzzing
sessions, and editing the generator in whatever order it finds most productive,
without being constrained to a fixed sequence of steps.

%% file: eval.tex
\section{Evaluation}
\label{sec:eval}

We evaluate \tool to explore how agent-synthesized generators compare to
human-written baselines, and to understand the impact of coverage guidance
and predicate-level feedback on generator quality.
Our evaluation is structured around four research questions:

\begin{itemize}
  \item[\textbf{RQ1}] Do agent-synthesized generators achieve higher coverage than human-written generators?
  \item[\textbf{RQ2}] Can agent-synthesized generators eliminate the need for coverage guidance and mutation-based fuzzing?
  \item[\textbf{RQ3}] Does fine-grained predicate feedback improve coverage of agent-synthesized generators?
  \item[\textbf{RQ4}] How does LLM predicate analysis compare to static dominator analysis?
\end{itemize}

\subsection{Experimental Setup}

\subsubsection{Techniques and Baselines}
\begin{figure*}[t]
  \centering
  \includegraphics[width=\linewidth]{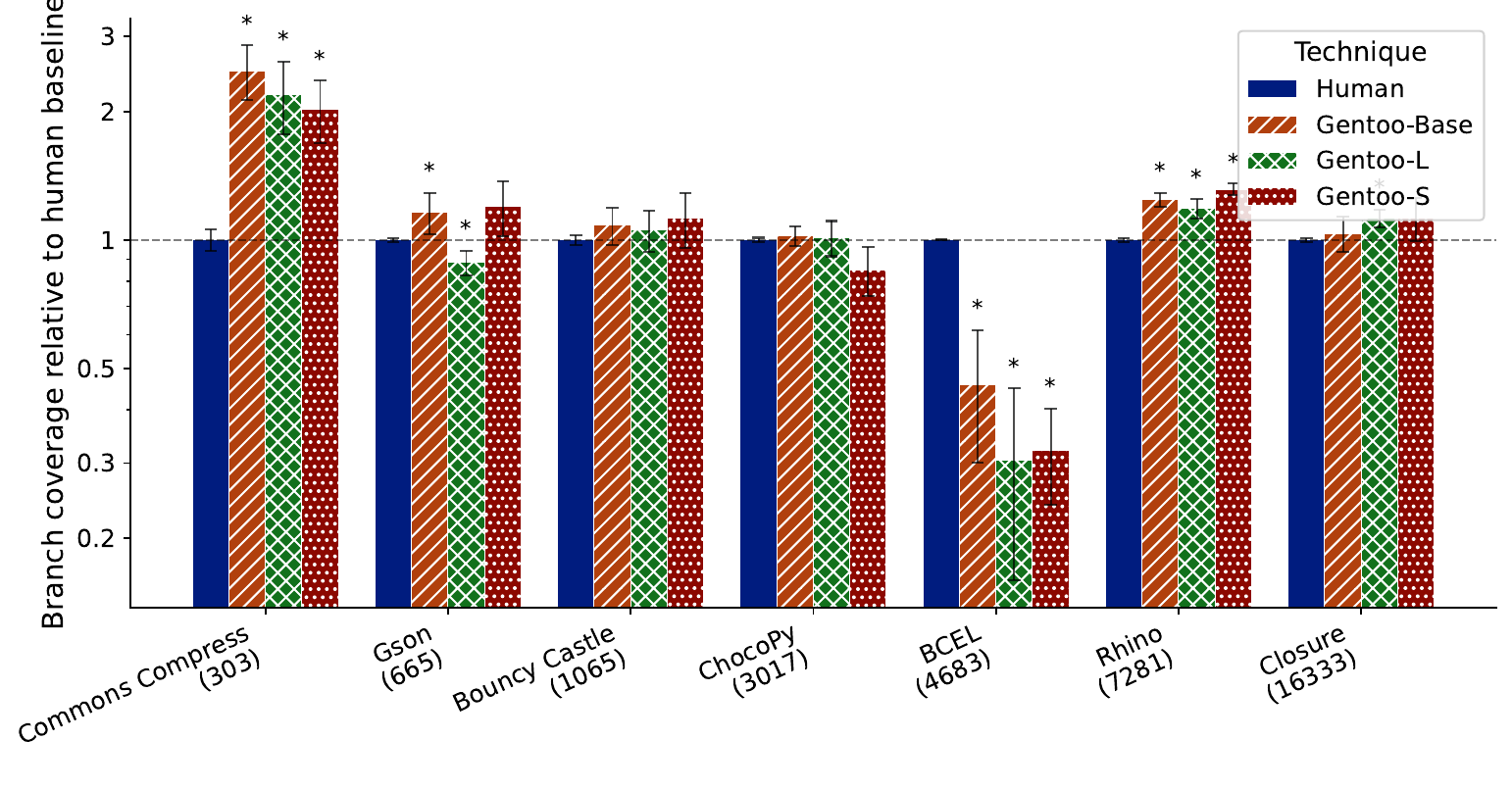}
  \caption{Branch coverage (mean $\pm$ std across five repetitions) for each
    technique and benchmark, normalized relative to the human-written generator
    baseline. Higher is better, * is statistically significance on a Mann-Whitney U test. In 4 out of 7 benchmarks, at least one agent-based approach significantly higher coverage.}
  \label{fig:rq1_coverage}
\end{figure*}

We evaluate all three \tool configurations (\toolStatic, \toolLLM, \toolBase)
against human-written generator baselines taken from prior
work~\cite{Padhye19-zest, mu2, Vikram21}.
For all configurations and baselines, we evaluate both coverage-guided
mutation-based fuzzing using the Zest algorithm and random input generation
(no mutations), to assess the contribution of the fuzzing engine independently
of generator quality.
The mean cost per synthesized generator is \$7.54 for \toolBase, \$12.17 for
\toolStatic, and \$12.84 for \toolLLM (using Claude Opus 4.5 via AWS Bedrock).

\subsubsection{Benchmarks}
We evaluate on six real-world Java library benchmarks, keeping fuzz targets
consistent with prior work~\cite{Padhye19-zest, Vikram21, mu2}:
\textbf{Apache BCEL} (Java bytecode parser),
\textbf{Bouncy Castle} (ASN.1/cryptographic structure parser),
\textbf{Google Closure} (JavaScript compiler),
\textbf{Apache Commons Compress} (BZip2 compressor/decompressor),
\textbf{Gson} (JSON parser), and
\textbf{Mozilla Rhino} (JavaScript interpreter).
These benchmarks span a range of input formats with hard structural
constraints, from recursive data formats (JSON, ASN.1) to language
compilers and interpreters, that are particularly challenging for
mutation-based fuzzers without structured input generators.

\subsubsection{Duration}
Each generator synthesis run is allotted a maximum of one hour of wall-clock time.
Each synthesized generator is then evaluated in a separate fuzzing campaign of
three hours.

\subsubsection{Repetitions}
All experiments are repeated five times with different random seeds, and we
report the mean and standard deviation across runs.

\subsubsection{Metrics}
We evaluate generators along two dimensions:
\begin{enumerate}
\item \textbf{Execution speed}: the average number of inputs executed per second
across the entire fuzzing campaign, reflecting the throughput cost of more complex generators.
\item \textbf{Branch coverage}: the number of branches covered during a fuzzing
campaign. We report coverage normalized relative to the human-written generator
baseline to enable comparison across benchmarks with different absolute coverage.
\end{enumerate}

\subsubsection{Scale}
In total, our evaluation comprises $3 \times 7 \times 5 = 105$ agent synthesis
runs across the three \tool configurations, 7 benchmarks, and 5 repetitions,
producing 101 synthesized generators (4 runs failed to produce a compilable
generator within the time budget).
Each synthesized generator and each human-written generator is evaluated in
both coverage-guided and random fuzzing modes, yielding 280 three-hour
fuzzing experiments in total.
The full evaluation required approximately 39 CPU-days of computation.

\subsection{RQ1: Coverage of synthesized generators}

\begin{framed}
\noindent\textit{Do agent-synthesized generators achieve higher coverage than human-written generators?}
\end{framed}

We evaluate branch coverage and execution throughput (as reported by JQF) across
all three \tool configurations and the human-written baselines.

\textbf{Branch coverage.}
Figure~\ref{fig:rq1_coverage} shows mean branch coverage normalized relative to
the human-written generator for each benchmark.
In 6 out of 7 benchmarks, at least one agent configuration matches or exceeds
the human baseline, with the best agent configuration ranging from $+2\%$
(ChocoPy) to $+148\%$ (Commons) above human coverage.
In 4 of those benchmarks (Closure, Commons, Gson, Rhino), at least one agent
configuration achieves statistically significantly higher coverage than the human
baseline (Mann-Whitney U, $p < 0.05$, marked with $*$ in Figure~\ref{fig:rq1_coverage}).
The exception is BCEL, where all agent configurations fall $54\%$ below the
human baseline; as discussed below, the human BCEL generator uses BCEL's own
high-level API to construct class files, giving it a structural advantage that
the agent does not discover.

For Rhino, \toolStatic achieves $31\%$ higher coverage than the human baseline,
reaching up to 9804 branches versus 7323 for the human generator.
The synthesized generator encodes domain-specific JavaScript constructs targeted
at Rhino's parser: \texttt{with} statements, labeled statements, hashbang lines,
and destructuring patterns, constructs the human generator does not produce,
each exercising distinct branches in Rhino's tokenizer and parser.
The agent discovers these constructs by reading Rhino's source and identifying
token types that correspond to rarely-reached parser states.

The ChocoPy result is surprising.
Agent coverage is only marginally above the human baseline ($+2\%$), despite
the agent targeting features related to semantic validity in many of the generators.
The human generator maintains a live list of declared class names and selects
a previously-declared class as the superclass for each new class definition,
naturally producing inheritance hierarchies by construction.
The agent does encode type-aware generation and propagates return types into
function bodies, but hard-codes a fixed set of class name pairs and generates
inheritance only in high-complexity code paths.
We hypothesize that the human generator's bounded design -- limiting identifier
pool size and AST depth -- increases the probability of type-consistent programs
by construction, without requiring explicit type tracking~\cite{Vikram21}.

The BCEL result is an instructive failure case.
All three agent configurations fall $54\%$ below the human baseline.
The human generator uses BCEL's own \texttt{org.apache.bcel.generic} API to
construct semantically valid class files from high-level abstractions, covering
nearly every JVM opcode via \texttt{InstructionFactory} and producing class files
that pass all three passes of BCEL's bytecode verifier.
The agent, by contrast, treats class file generation as a raw byte-stream encoding
problem, manually writing magic numbers and constant pool entries, producing
inputs that frequently fail early in the parser with a format exception.
None of the agent configurations discover the high-level construction API despite
it being present in the library source.

\mybox{\textbf{Finding 1:} Agent-synthesized generators achieve statistically significantly higher branch coverage than human-written generators on 4 of 7 benchmarks. The BCEL exception highlights a gap in agent strategy: failing to discover and leverage the library's own high-level construction API.}

\begin{table}[t]
  \footnotesize
  \centering
  \caption{Mean execution speed (inputs/sec $\pm$ std across five repetitions)
    for each technique and benchmark, reflecting the throughput cost of
    more complex generators.}
  \label{tab:rq1_speed}
  \input{figs/rq1_speed_table}
\end{table}

\textbf{Execution speed.}
Agent-synthesized generators tend to produce more complex inputs than the human
baseline, resulting in lower throughput.
Across non-BCEL benchmarks, agent configurations are $81\%$ slower to $79\%$
faster than the human generator, with a median slowdown of $32\%$.
Rhino shows the highest variance (26--149 inputs/sec for \toolLLM): some
repetitions synthesize a generator with a shallow recursion-depth limit and no
guard on nested block generation, producing accidentally deep JavaScript programs
that are slow to interpret.
BCEL is the outlier in the opposite direction: agent generators run at $\sim$590
inputs/sec versus $\sim$92 for the human generator, since the human generator
invokes BCEL's own class-building API to construct real class file objects,
which is considerably more expensive than the agent's raw byte-writing approach.

\subsection{RQ2: Impact of Coverage Guidance}

\begin{framed}
\noindent\textit{Can agent-synthesized generators eliminate the need for coverage guidance and mutation-based fuzzing?}
\end{framed}

\begin{figure}[t]
  \centering
  \includegraphics[width=\linewidth]{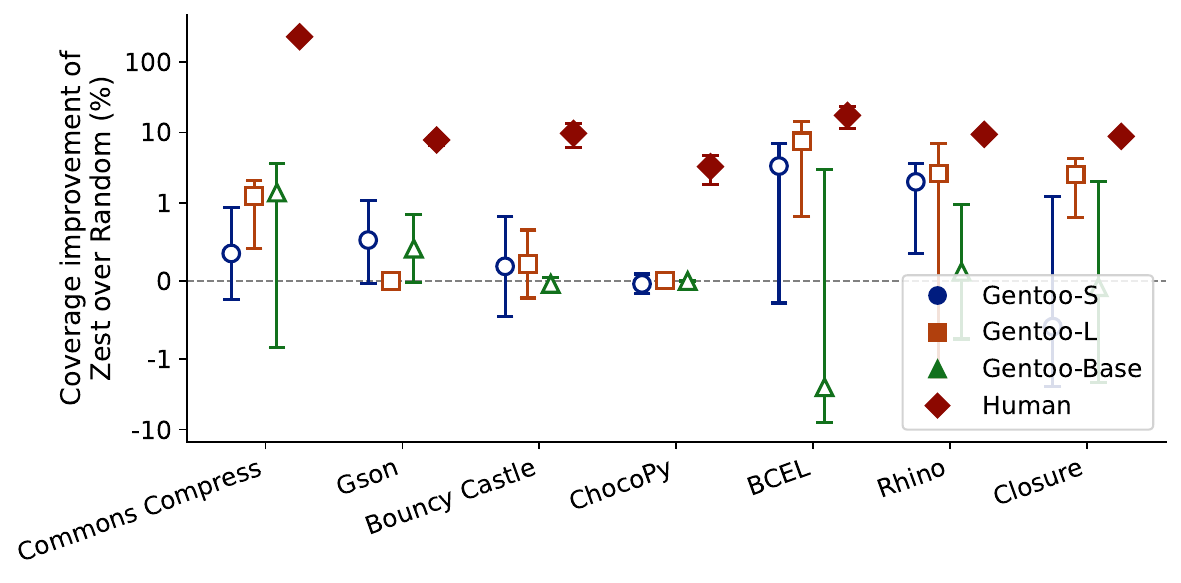}
  \caption{Percentage improvement in final branch coverage of coverage-guided
    fuzzing (Zest) over random sampling for each technique and benchmark.
    Dots above zero indicate that Zest helps; dots below zero indicate it hurts.
    Filled markers denote a statistically significant difference
    ($p < 0.05$, MWU); hollow markers are not significant.}
  \label{fig:rq4}
\end{figure}

We evaluate each synthesized and human-written generator under two conditions:
coverage-guided mutation-based fuzzing using the Zest algorithm, and random
input generation with no mutation feedback. In both conditions, branch coverage
is tracked over the three-hour campaign.
Figure~\ref{fig:rq4} shows the percentage improvement in final branch coverage
of coverage-guided fuzzing over random sampling for \toolStatic, \toolLLM, and
\toolBase across all benchmarks.

For all three agent configurations, coverage guidance never produces a
statistically significant improvement over random input generation on any
benchmark, with average differences of $-0.1\%$, $+0.8\%$, and $+2.0\%$ for
\toolBase, \toolStatic, and \toolLLM respectively.
By contrast, for the human-written generators, coverage guidance yields a
statistically significant improvement on all 7 benchmarks, with an average
improvement of $+40\%$ (ranging from $+3\%$ on ChocoPy to $+229\%$ on Commons).

This contrast has a natural explanation.
Many of the human-written generators are relatively simple -- naive byte
sequences, grammar samplers, or shallow random programs -- for which mutation
and coverage feedback can meaningfully guide exploration toward structurally
valid inputs.
The Commons benchmark illustrates this directly: the human generator produces
purely random byte sequences, so coverage guidance yields a $229\%$ improvement
as the fuzzer evolves the corpus toward inputs that satisfy the BZip2 stream
header (\texttt{BZh}), block magic bytes, and Huffman table structure needed to
reach the decompressor's inner logic.
The agent-synthesized generator, by contrast, explicitly encodes this structure:
it generates the \texttt{BZh} header, block magic, CRC fields, randomised flags,
and Huffman coding tables from scratch, producing valid BZip2 streams on every
invocation.
When a generator already encodes such domain-specific structure, coverage-guided
mutation has little room to improve over purely random sampling from the same
generator.

Figure~\ref{fig:chocopy-snippet} shows an example from the agent-synthesized
ChocoPy generator: methods are generated with typed \texttt{self} parameters,
declared return types, and return values that are type-consistent by construction. 
Notably, the generator always returns a literal (e.g., an integer or boolean
constant) rather than a more complex expression, trading input diversity for
guaranteed type correctness -- a conservative strategy focused on semantic
validity that makes coverage guidance and mutation strategies less effective.

\begin{figure}[t]
\begin{minted}[fontsize=\scriptsize,frame=single,breaklines]{java}
// (Simplified) self is typed with the enclosing class name;
// return value is a type-consistent literal.
private void generateMethod(String className, String methodName) {
  String returnType = chooseType(random);
  sb.append("def ").append(methodName)
    .append("(self: ").append(className).append(") -> ")
    .append(returnType).append(":\n");
  sb.append("    return ")
    .append(generateLiteralForType(random, returnType)).append("\n");
}

private String generateLiteralForType(SourceOfRandomness random, String type) {
  switch (type) {
    case "int":  return String.valueOf(random.nextInt(-100, 100));
    case "bool": return random.nextBoolean() ? "True" : "False";
    case "str":  return "\"" + generateSimpleString(random) + "\"";
    default:     return "None";
  }
}
\end{minted}
\caption{Simplified excerpt from a \toolLLM agent-synthesized ChocoPy generator.
  The generator encodes ChocoPy type semantics directly: \texttt{self} is
  typed with the enclosing class name and return values are constrained to
  match the method's declared return type. By always returning a type-consistent literal, the generator trades input
  diversity for guaranteed well-typedness.}
\label{fig:chocopy-snippet}
\end{figure}

\mybox{\textbf{Finding 2:} Coverage guidance and mutation strategies are never statistically significantly beneficial for agent-synthesized generators, but are significant for all human-written generators. Structural and semantic logic encoded in agent generators makes coverage guidance and mutation largely unnecessary.}

\subsection{RQ3: Impact of Predicate Feedback}

\begin{framed}
\noindent\textit{Does fine-grained predicate feedback improve coverage of agent-synthesized generators?}
\end{framed}

Comparing final branch coverage across all three configurations
(Figure~\ref{fig:rq1_coverage}), \toolBase leads on 3 benchmarks, \toolStatic
on 3, and \toolLLM on 1, with differences often within the variance across
repetitions.
This suggests that a well-prompted baseline agent can synthesize comparably
strong generators without predicate guidance.

\begin{figure*}[t]
  \centering
  \begin{subfigure}[t]{0.24\linewidth}
    \includegraphics[width=\linewidth]{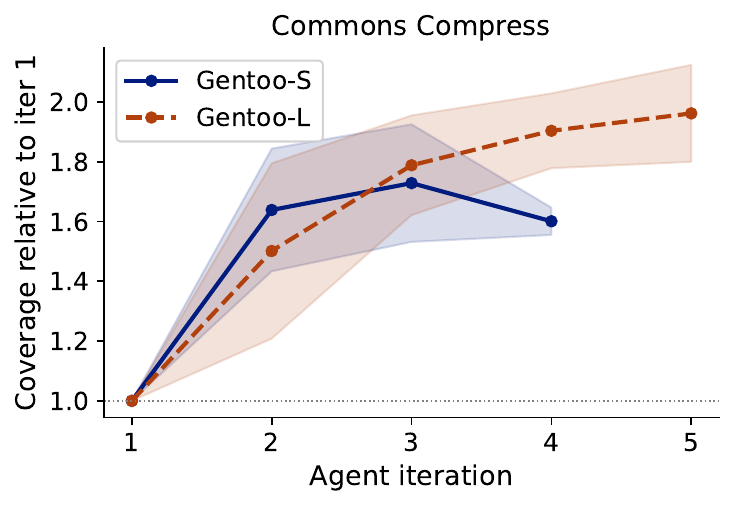}
  \end{subfigure}\hfill
  \begin{subfigure}[t]{0.24\linewidth}
    \includegraphics[width=\linewidth]{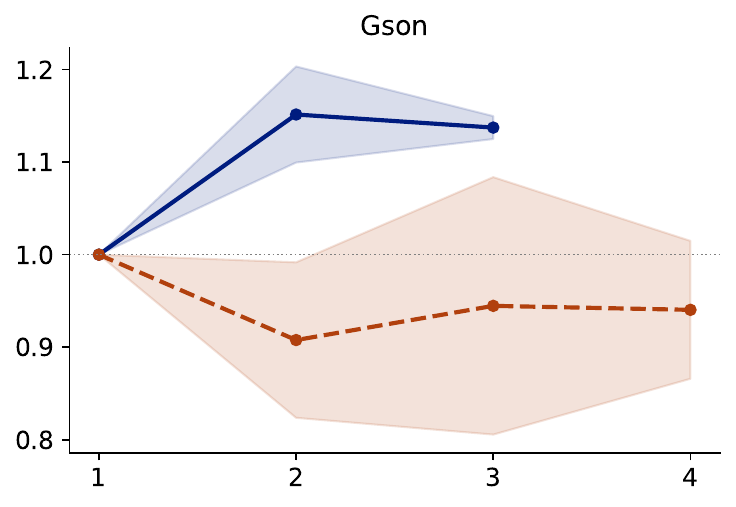}
  \end{subfigure}\hfill
  \begin{subfigure}[t]{0.24\linewidth}
    \includegraphics[width=\linewidth]{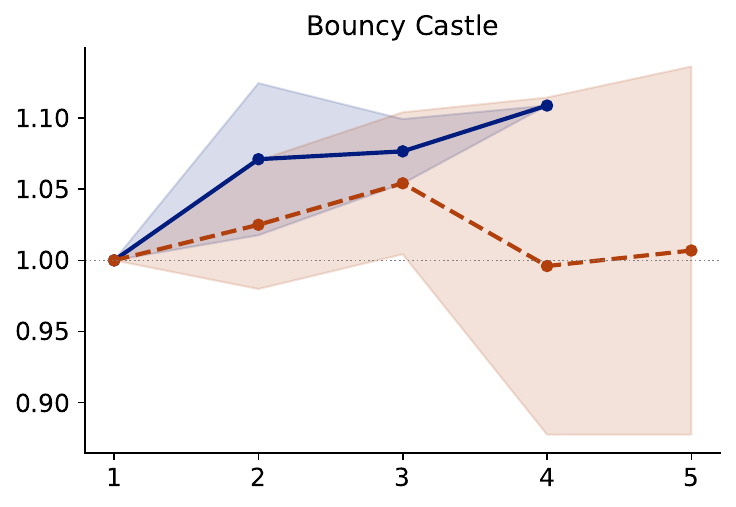}
  \end{subfigure}\hfill
  \begin{subfigure}[t]{0.24\linewidth}
    \includegraphics[width=\linewidth]{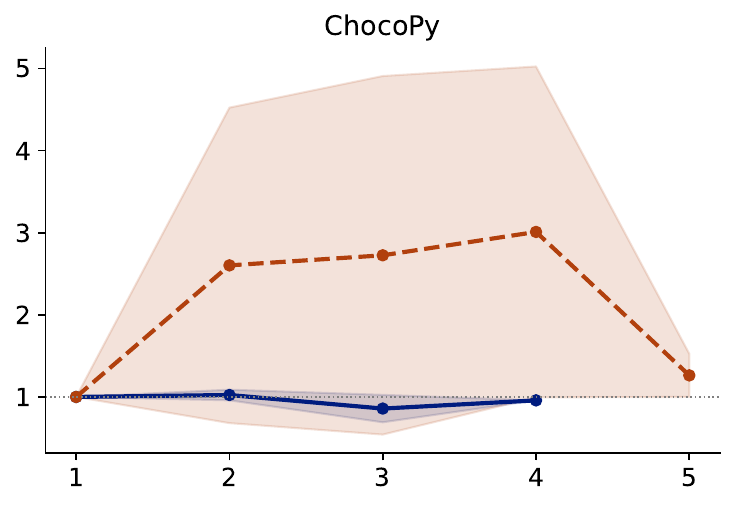}
  \end{subfigure}

  \medskip

  \begin{center}
  \begin{subfigure}[t]{0.24\linewidth}
    \includegraphics[width=\linewidth]{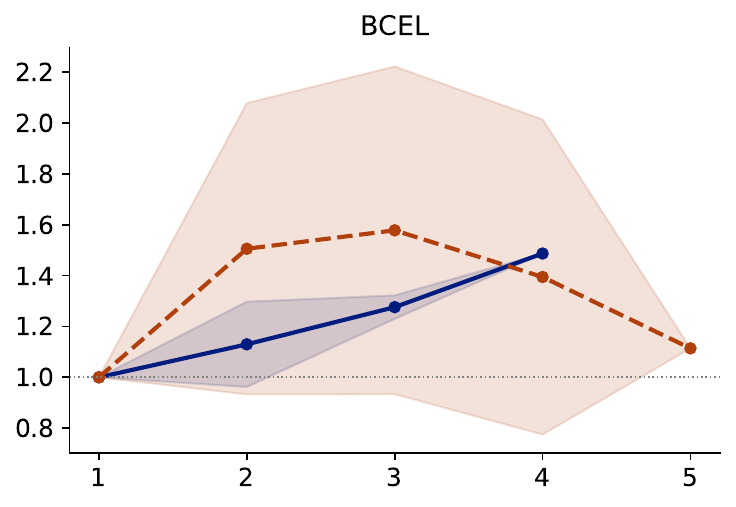}
  \end{subfigure}\hspace{0.013\linewidth}
  \begin{subfigure}[t]{0.24\linewidth}
    \includegraphics[width=\linewidth]{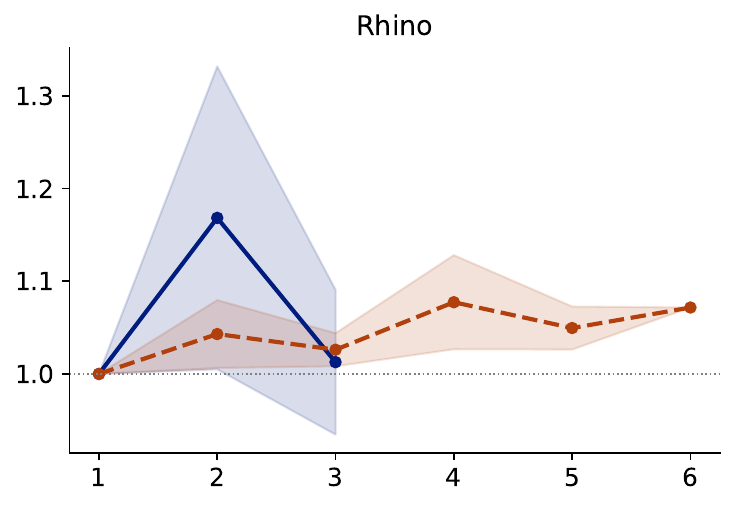}
  \end{subfigure}\hspace{0.013\linewidth}
  \begin{subfigure}[t]{0.24\linewidth}
    \includegraphics[width=\linewidth]{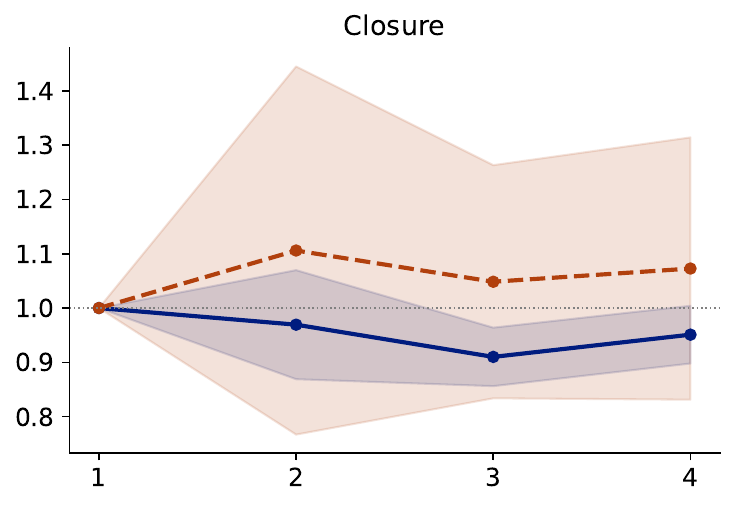}
  \end{subfigure}
  \end{center}

  \caption{Branch coverage relative to iteration~1 over agent iterations
    for \toolStatic (solid) and \toolLLM (dashed).
    A value of 1.0 means no change from the first iteration; higher is better.
    Shading shows $\pm1$ std across five repetitions.}
  \label{fig:rq2-cov-iters}
\end{figure*}

To understand whether predicate feedback contributes within the synthesis
process itself, we also examine how coverage evolves across agent iterations.
For \toolStatic and \toolLLM, each time the agent invokes the fuzzing
campaign to test its current generator, we checkpoint both the generator and
the fuzzing results.
We then measure branch coverage at each checkpoint relative to the first
iteration to isolate the contribution of iterative refinement.
Figure~\ref{fig:rq2-cov-iters} shows this per-iteration coverage ratio
for \toolStatic (solid) and \toolLLM (dashed) across all benchmarks.

Coverage improves in 74\% of runs for both \toolStatic and \toolLLM.
\toolStatic achieves a mean improvement of $+18\%$ (median $+12\%$) over the
first iteration, while \toolLLM achieves a mean of $+57\%$ (median $+11\%$),
with the higher mean driven by large gains on ChocoPy and BCEL where the initial
generator is weak and predicate feedback guides the agent to add substantial new
generation logic.
Most benchmarks show a monotone improvement trend, with coverage plateauing
after two to three iterations as the agent exhausts the changes suggested by
the predicate feedback.

We do not observe monotonic increases in coverage across iterations on all
benchmarks.
In one such case, the \toolStatic generator for Closure regresses in 3 of 5
repetitions: after observing zero coverage on the highest-dominance predicates,
the agent attempts to increase the compilation success rate reported by Zest,
simplifying the generator to produce shorter and more conservative JavaScript,
which reduces input diversity and lowers coverage.
This illustrates a limitation: predicate feedback may lead the agent to optimize
in the wrong direction when the signal is ambiguous.

The Commons Compress benchmark illustrates a successful refinement.
Between iteration~1 and iteration~2, \toolStatic replaces four approximate
strategy methods with six bit-precise methods encoding specific BZip2 structures,
and critically bounds the \texttt{origPtr} field by the actual data size,
fixing an \texttt{IOException} that had blocked the decompressor's inner logic.
Coverage improves by $88\%$ in a single refinement step.

%
%

\mybox{\textbf{Finding 3:} Predicate feedback improves generators in 74\% of runs, but final branch coverage is broadly comparable across all three configurations, suggesting predicate feedback is beneficial but not decisive compared to a well-prompted baseline agent.}

\subsection{RQ4: LLM vs.\ Static Predicate Analysis}

\begin{framed}
\noindent\textit{How does LLM predicate analysis compare to static dominator analysis?}
\end{framed}

As shown in Figure~\ref{fig:rq1_coverage}, \toolStatic and \toolLLM achieve
broadly comparable final coverage, with \toolStatic leading on 3 benchmarks
and \toolLLM on 1.
Despite this similarity in outcome, the two approaches identify different
predicates and reason about them differently.

Static dominator analysis ranks predicates purely by how many CFG nodes they
structurally dominate, without any notion of how hard it is for an input
generator to satisfy each branch.
The predicate coverage report provides a runtime measurement of this hardness
after execution, but not before.
The LLM, by contrast, reads the source code and reasons about both dominance
and semantic hardness before execution of the generators, filtering out
high-dominance predicates that are easy to satisfy and focusing on those
that require non-trivial input structure.

The Rhino benchmark illustrates this clearly.
Static analysis ranks token-stream predicates at the top (dominance scores
628--651), but these are trivially satisfied -- over 10{,}000 inputs hit all
branches since any sequence of characters exercises different token types.
\toolLLM instead targets parser-level predicates for statements, expressions,
and function definitions (estimated dominance 100--200), correctly reasoning
that reaching the parser requires syntactically structured JavaScript programs
rather than arbitrary token sequences.

\mybox{\textbf{Finding 4:} \toolStatic and \toolLLM achieve comparable final coverage despite targeting different predicates. Static analysis ranks by structural dominance alone, while the LLM additionally reasons about semantic hardness -- sometimes identifying hard-to-reach predicates that static analysis deprioritizes.}

%% file: figs/rq1_speed_table.tex
\begin{tabular}{lrrrr}
\toprule
Benchmark & Human & Gentoo-Base & Gentoo-L & Gentoo-S \\
\midrule
bcel & 91.5 $\pm$ 21.7 & 594.9 $\pm$ 3.4 & 583.9 $\pm$ 27.3 & 593.7 $\pm$ 3.4 \\
bouncycastle & 594.5 $\pm$ 3.4 & 594.6 $\pm$ 3.5 & 593.4 $\pm$ 3.6 & 593.6 $\pm$ 3.5 \\
chocopy & 250.6 $\pm$ 10.0 & 104.1 $\pm$ 12.8 & 124.1 $\pm$ 45.9 & 108.2 $\pm$ 57.5 \\
closure & 36.2 $\pm$ 2.4 & 19.1 $\pm$ 6.4 & 40.9 $\pm$ 12.8 & 17.4 $\pm$ 11.1 \\
commons & 126.9 $\pm$ 64.2 & 29.3 $\pm$ 22.2 & 84.2 $\pm$ 63.4 & 23.5 $\pm$ 20.0 \\
gson & 595.8 $\pm$ 2.6 & 409.6 $\pm$ 145.2 & 485.5 $\pm$ 119.6 & 488.5 $\pm$ 146.8 \\
rhino & 83.3 $\pm$ 10.9 & 112.4 $\pm$ 42.6 & 56.6 $\pm$ 25.0 & 148.8 $\pm$ 66.1 \\
\bottomrule
\end{tabular}

%% file: related.tex
\section{Related Work}
\label{sec:related}
There has recently been a large body of work to use large language models for software testing tasks~\cite{wang2024openhands, hou2023large, schafer2023empirical}. 

\paragraph{Automatic Fuzz Driver Synthesis} The problem of \emph{fuzz harness} or \emph{fuzz driver} generation bears similarity to our input generator synthesis task~\cite{babic2019fudge,ispoglou2020fuzzgen}. In fuzz driver generation, the goal is to take unstructured byte data provided by the fuzz tester, and use it to exercise the program under test in a meaningful manner. UTopia~\cite{jeong2023utopia} extracts fuzz drivers from unit tests, noting that some unit test assertions (e.g., checking null pointers) must be preserved to maintain property validity. LLMs have also been applied to fuzz driver generation~\cite{zhang2023understanding}.
PromptFuzz~\cite{lyu2023prompt} runs harnesses during synthesis and uses coverage
information to select which functions to include in the next harness prompt.
CKGFuzzer~\cite{xu2024ckgfuzzer} introduces a code knowledge graph for better
library understanding, and PromeFuzz~\cite{liu2025promefuzz} employs
retrieval-augmented generation and an LLM-based bug analyzer.
OSS-Fuzz-Gen~\cite{ossfuzz23} has successfully landed new fuzz harnesses in
real open-source projects via Google's OSS-Fuzz infrastructure.
Stitch~\cite{green2026stitch} takes a different approach, encoding API usage
constraints as composable blocks that a fuzzer dynamically assembles at runtime,
with a static type system governing object flow between blocks and a
dynamically-checked typestate tracking semantic constraints such as object state
dependencies and cross-function preconditions.

\paragraph{Large Language Models for Input Generation}
Recently, researchers have developed methods of providing specific documentation as context to the LLM for the purpose of generating test inputs that exercise specific behavior. Fuzz4All~\cite{xia2024fuzz4all} provides language standards to develop and \textit{autoprompting} workflow to prompt LLMs to act as input generation and mutation engines and fuzz various compilers. DiffSpec~\cite{rao2024diffspec} compiles historical bug data and typing rules from documentation to provide to LLMs in order to differentially test the \code{wasm} validator. Ackermann et al.~\cite{ackerman2023large} utilize LLMs to examine natural language format specifications to generate strong seeds for a mutation-based fuzzing. Meng et al.~\cite{meng2024large} develop ChatAFL to enhance protocol fuzzing by enriching initial seeds with specific LLM message outputs. FuzzGPT~\cite{deng2023large} and WhiteFox~\cite{yang2023whitefox} prompt LLMs to synthesize Python programs to test deep-learning libraries and compilers. EvalPlus augments the HumanEval unit testing dataset by prompting the LLM to produce seed inputs and performing further type-based mutation~\cite{liu2024your}. While these approaches have been successful at improving upon existing fuzzing baselines, Jiang et al.~\cite{jiang2024when} detail limitations including insufficient input diversity and limited validity.

LLM-based test generation is a closely related line of work.
LLMs have been applied to generate unit tests~\cite{bareiss2022pradeltest,schafer2023testpilot,rao2023cat,alshahwan2024automated},
co-generate code and tests interactively~\cite{lahiri2022interactive},
and aid search-based unit test generation~\cite{lemieux2023codamosa}.
TestPilot~\cite{schafer2023testpilot} uses automated prompt refinement,
enriching prompts with documentation, signatures, or error messages when a
generated test fails.
LLMs have also been applied to property-based test generation~\cite{vikram2023can, maaz2025agentic},
which shares with our work the goal of choosing input generation strategies that produce inputs rather than synthesizing individual test cases.

\paragraph{LLM-Synthesized Input Generators}
ELFuzz~\cite{chen2025elfuzz} and G\textsuperscript{2}Fuzz~\cite{zhang2025low} are
the most closely related works to \tool.
ELFuzz uses an LLM-driven evolution loop to iteratively mutate generator code,
guided by a coverage-range lattice (``fuzzer space'') that enables fine-grained
comparison of which code regions each fuzzer variant covers.
G\textsuperscript{2}Fuzz synthesizes Python generators for non-textual binary
formats and re-invokes the LLM when AFL++ stalls.
Both approaches treat the LLM as a stateless code mutator: given a generator,
produce a better one.
\tool differs in two key ways: it deploys an autonomous agent with terminal
access that can compile, execute, and debug generators over multiple reasoning
steps; and it grounds refinement in predicate-level feedback, directly telling the agent which high-impact
program predicates its generator fails to satisfy.
Neither ELFuzz nor G\textsuperscript{2}Fuzz focus their evaluation on agentic
synthesis methods or investigate the impact of coverage guidance on the generators; both evaluate against traditional mutation-based
fuzzers.

%% file: threats.tex
\section{Threats to Validity}
\label{sec:threats}

\paragraph{Construct Validity.}
We use branch coverage as the primary metric for evaluating generator quality.
Branch coverage is a necessary but not sufficient condition for finding bugs:
a generator that covers more code reaches more potential bug sites, but
coverage alone does not guarantee fault detection.
A second construct threat is the use of per-iteration coverage improvement as
a proxy for the utility of predicate feedback.
An agent may improve coverage between iterations for reasons unrelated to
predicate feedback, for example by restructuring generation logic based on
inspecting generated inputs, reading source code more carefully, or simply
exploring a different region of the generator design space.
We cannot fully attribute these improvements to predicate
feedback specifically.

\paragraph{Internal Validity.}
A threat to internal validity is the non-determinism inherent in both LLM
generation and coverage-guided fuzzing.
LLM outputs vary across runs even for the same prompt, and fuzzing campaigns
are randomized.
We mitigate this by running each configuration five times and reporting
aggregate statistics.
A second threat is that the LLM agent we use (Claude) may have been exposed to
the source code of the libraries under test during pretraining, potentially
giving it an advantage over a truly black-box baseline.
We cannot fully rule out this possibility, but we note that any memorized
knowledge must be translated into executable generation logic to be useful.
Upon manually investigating a sample of the agent-synthesized generators, we note
that they do not share any exact similarity in generation logic compared to the
human-written generators.
Finally, our evaluation measures coverage at the end of a fixed-time fuzzing
campaign; the relative ordering of configurations could differ under longer
or shorter time budgets.

\paragraph{External Validity.}
Our evaluation targets Java libraries fuzzed via the JQF framework.
Our conclusions may not generalize to programs written in other languages or
fuzzed with other tools such as AFL~\cite{Afl} or libFuzzer~\cite{libfuzzer},
where different generator interfaces and instrumentation mechanisms apply.
Additionally, our benchmark suite covers 7 libraries across diverse input
formats; programs with very large or dynamically loaded codebases may be
harder to analyze statically.
Our evaluation uses Claude 4.5 Opus as the underlying model with the Claude Agent SDK 
scaffolding; results may differ with other models, agent frameworks, or
prompt strategies, and we cannot claim that our findings generalize to
all LLM-based synthesis approaches.

%% file: discussion.tex
\section{Discussion and Conclusion}
\label{sec:discussion}

\paragraph{Generators are all you need.}
Our results show that coverage-guided mutation contributes only marginally
on top of agent-synthesized generators, and never significantly so.
When a generator encodes the target's structural and semantic constraints
directly, the fuzzing loop becomes a simple execution harness.
This suggests a shift in focus from fuzzer engineering to generator synthesis:
investing in better generators may be more productive than investing in smarter
mutation strategies or power schedules.

\paragraph{Agents and application-specific analysis.}
Our results offer a positive example of a broader trend: coding agents
constructing application-specific analyses tailored to a given target, rather
than applying a fixed general-purpose tool uniformly.
For each library, the agent reads the source, reasons about which input
constraints matter for that specific target, and writes generation logic
that encodes those constraints directly.
This contrasts with traditional approaches that apply the same fuzzer,
symbolic executor, or static analyzer to every target and rely on the tool
to bridge the gap to application-specific behavior.
We believe this paradigm, agents as constructors of bespoke per-target
analyses, is broadly applicable beyond input generation, and represents
a promising direction for automating software engineering tasks that have
historically resisted general-purpose tooling.

\paragraph{The future of agents and generator-based fuzzing.}
Today's coding agents primarily choose unit testing as their default behavior as a 
means of validating correctness. We hope to see a shift toward higher-scale automated testing techniques such as generator-based fuzzing, where the agent's ability to synthesize domain-specific input generators on
the fly unlocks a different scale of test-input exploration. As agents become more capable, the barrier to building such generators will
continue to fall, potentially making generator-based fuzzing a practical first-class
option for automated software validation.

\section{Data Availability}
\label{sec:data}
The source code of \tool, the synthesized input generators for all benchmarks, execution results, and scripts from our evaluation are available on Zenodo at
\url{https://doi.org/10.5281/zenodo.19138370}.

\section{Acknowledgments}
\label{sec:acks}
This work was supported in part by the National Science Foundation grant CCF-2120955 and an Amazon Research Award.

%% file: paper.bbl

\begin{thebibliography}{68}


\ifx \showCODEN    \undefined \def \showCODEN     #1{\unskip}     \fi
\ifx \showDOI      \undefined \def \showDOI       #1{#1}\fi
\ifx \showISBNx    \undefined \def \showISBNx     #1{\unskip}     \fi
\ifx \showISBNxiii \undefined \def \showISBNxiii  #1{\unskip}     \fi
\ifx \showISSN     \undefined \def \showISSN      #1{\unskip}     \fi
\ifx \showLCCN     \undefined \def \showLCCN      #1{\unskip}     \fi
\ifx \shownote     \undefined \def \shownote      #1{#1}          \fi
\ifx \showarticletitle \undefined \def \showarticletitle #1{#1}   \fi
\ifx \showURL      \undefined \def \showURL       {\relax}        \fi
\providecommand\bibfield[2]{#2}
\providecommand\bibinfo[2]{#2}
\providecommand\natexlab[1]{#1}
\providecommand\showeprint[2][]{arXiv:#2}

\bibitem[wal({[n.\,d.]})]%
        {wala}
 \bibinfo{year}{[n.\,d.]}\natexlab{}.
\newblock \bibinfo{title}{{WALA}: {T.J.} Watson Libraries for Analysis}.
\newblock \bibinfo{howpublished}{\url{https://github.com/wala/wala}}.
\newblock
\newblock
\shownote{Accessed 2025}.


\bibitem[jun(nd)]%
        {junit-quickcheck-implied}
 \bibinfo{year}{[n.\,d.]}\natexlab{}.
\newblock \bibinfo{title}{junit-quickcheck-generator}.
\newblock
  \bibinfo{howpublished}{\url{https://pholser.github.io/junit-quickcheck/site/1.0/usage/other-types.html}}.
\newblock
\newblock
\shownote{Accessed: 2024-10-31}.


\bibitem[lib(nd)]%
        {libfuzzer}
 \bibinfo{year}{[n.\,d.]}\natexlab{}.
\newblock \bibinfo{title}{libFuzzer – a library for coverage-guided fuzz
  testing.}
\newblock \bibinfo{howpublished}{\url{https://llvm.org/docs/LibFuzzer.html}}.
\newblock
\newblock
\shownote{Accessed: 2021-08-31}.


\bibitem[Ackerman and Cybenko(2023)]%
        {ackerman2023large}
\bibfield{author}{\bibinfo{person}{Joshua Ackerman} {and}
  \bibinfo{person}{George Cybenko}.} \bibinfo{year}{2023}\natexlab{}.
\newblock \showarticletitle{Large language models for fuzzing parsers
  (registered report)}. In \bibinfo{booktitle}{\emph{Proceedings of the 2nd
  International Fuzzing Workshop}}. \bibinfo{pages}{31--38}.
\newblock


\bibitem[Alshahwan et~al\mbox{.}(2024)]%
        {alshahwan2024automated}
\bibfield{author}{\bibinfo{person}{Nadia Alshahwan}, \bibinfo{person}{Jubin
  Chheda}, \bibinfo{person}{Anastasia Finegenova}, \bibinfo{person}{Beliz
  Gokkaya}, \bibinfo{person}{Mark Harman}, \bibinfo{person}{Inna Harper},
  \bibinfo{person}{Alexandru Marginean}, \bibinfo{person}{Shubho Sengupta},
  {and} \bibinfo{person}{Eddy Wang}.} \bibinfo{year}{2024}\natexlab{}.
\newblock \showarticletitle{Automated unit test improvement using large
  language models at meta}.
\newblock \bibinfo{journal}{\emph{arXiv preprint arXiv:2402.09171}}
  (\bibinfo{year}{2024}).
\newblock


\bibitem[Anthropic({[n.\,d.]})]%
        {claude-agent-sdk}
\bibfield{author}{\bibinfo{person}{Anthropic}.}
  \bibinfo{year}{[n.\,d.]}\natexlab{}.
\newblock \bibinfo{title}{Claude Agent {SDK}}.
\newblock
  \bibinfo{howpublished}{\url{https://platform.claude.com/docs/en/agent-sdk/overview}}.
\newblock
\newblock
\shownote{Accessed 2025}.


\bibitem[Aschermann et~al\mbox{.}(2019)]%
        {Aschermann19-redqueen}
\bibfield{author}{\bibinfo{person}{Cornelius Aschermann},
  \bibinfo{person}{Sergej Schumilo}, \bibinfo{person}{Tim Blazytko},
  \bibinfo{person}{Robert Gawlik}, {and} \bibinfo{person}{Thorsten Holz}.}
  \bibinfo{year}{2019}\natexlab{}.
\newblock \showarticletitle{REDQUEEN: Fuzzing with Input-to-State
  Correspondence.}. In \bibinfo{booktitle}{\emph{NDSS}},
  Vol.~\bibinfo{volume}{19}. \bibinfo{pages}{1--15}.
\newblock
\urldef\tempurl%
\url{https://doi.org/10.14722/ndss.2019.23371}
\showDOI{\tempurl}


\bibitem[Ba and Rigger(2023)]%
        {ba2023testing}
\bibfield{author}{\bibinfo{person}{Jinsheng Ba} {and} \bibinfo{person}{Manuel
  Rigger}.} \bibinfo{year}{2023}\natexlab{}.
\newblock \showarticletitle{Testing database engines via query plan guidance}.
  In \bibinfo{booktitle}{\emph{2023 IEEE/ACM 45th International Conference on
  Software Engineering (ICSE)}}. IEEE, \bibinfo{pages}{2060--2071}.
\newblock


\bibitem[Babi{\'c} et~al\mbox{.}(2019)]%
        {babic2019fudge}
\bibfield{author}{\bibinfo{person}{Domagoj Babi{\'c}}, \bibinfo{person}{Stefan
  Bucur}, \bibinfo{person}{Yaohui Chen}, \bibinfo{person}{Franjo
  Ivan{\v{c}}i{\'c}}, \bibinfo{person}{Tim King}, \bibinfo{person}{Markus
  Kusano}, \bibinfo{person}{Caroline Lemieux}, \bibinfo{person}{L{\'a}szl{\'o}
  Szekeres}, {and} \bibinfo{person}{Wei Wang}.}
  \bibinfo{year}{2019}\natexlab{}.
\newblock \showarticletitle{Fudge: fuzz driver generation at scale}. In
  \bibinfo{booktitle}{\emph{Proceedings of the 2019 27th ACM Joint Meeting on
  European Software Engineering Conference and Symposium on the Foundations of
  Software Engineering}}. \bibinfo{pages}{975--985}.
\newblock


\bibitem[Barei{\ss} et~al\mbox{.}(2022)]%
        {bareiss2022pradeltest}
\bibfield{author}{\bibinfo{person}{Patrick Barei{\ss}},
  \bibinfo{person}{Beatriz Souza}, \bibinfo{person}{Marcelo d'Amorim}, {and}
  \bibinfo{person}{Michael Pradel}.} \bibinfo{year}{2022}\natexlab{}.
\newblock \showarticletitle{Code generation tools (almost) for free? a study of
  few-shot, pre-trained language models on code}.
\newblock \bibinfo{journal}{\emph{arXiv preprint arXiv:2206.01335}}
  (\bibinfo{year}{2022}).
\newblock


\bibitem[B{\"o}hme et~al\mbox{.}(2016)]%
        {Bohme16}
\bibfield{author}{\bibinfo{person}{Marcel B{\"o}hme},
  \bibinfo{person}{Van-Thuan Pham}, {and} \bibinfo{person}{Abhik
  Roychoudhury}.} \bibinfo{year}{2016}\natexlab{}.
\newblock \showarticletitle{Coverage-based Greybox Fuzzing as Markov Chain}. In
  \bibinfo{booktitle}{\emph{Proceedings of the 2016 ACM SIGSAC Conference on
  Computer and Communications Security}} \emph{(\bibinfo{series}{CCS})}.
  \bibinfo{pages}{1032--1043}.
\newblock
\urldef\tempurl%
\url{https://doi.org/10.1145/2976749.2978428}
\showDOI{\tempurl}


\bibitem[B{\"o}hme et~al\mbox{.}(2017)]%
        {aflfast}
\bibfield{author}{\bibinfo{person}{Marcel B{\"o}hme},
  \bibinfo{person}{Van-Thuan Pham}, {and} \bibinfo{person}{Abhik
  Roychoudhury}.} \bibinfo{year}{2017}\natexlab{}.
\newblock \showarticletitle{Coverage-based greybox fuzzing as markov chain}.
\newblock \bibinfo{journal}{\emph{IEEE Transactions on Software Engineering}}
  \bibinfo{volume}{45}, \bibinfo{number}{5} (\bibinfo{year}{2017}),
  \bibinfo{pages}{489--506}.
\newblock


\bibitem[B{\"o}hme et~al\mbox{.}(2022)]%
        {Bohme22}
\bibfield{author}{\bibinfo{person}{Marcel B{\"o}hme},
  \bibinfo{person}{L{\'a}szl{\'o} Szekeres}, {and} \bibinfo{person}{Jonathan
  Metzman}.} \bibinfo{year}{2022}\natexlab{}.
\newblock \showarticletitle{On the Reliability of Coverage-Based Fuzzer
  Benchmarking}. In \bibinfo{booktitle}{\emph{44th {IEEE/ACM} International
  Conference on Software Engineering}} \emph{(\bibinfo{series}{ICSE'22})}.
\newblock
\urldef\tempurl%
\url{https://doi.org/10.1145/3510003.3510230}
\showDOI{\tempurl}
\newblock
\shownote{to appear}.


\bibitem[Cha et~al\mbox{.}(2015)]%
        {Cha15}
\bibfield{author}{\bibinfo{person}{Sang~Kil Cha}, \bibinfo{person}{Maverick
  Woo}, {and} \bibinfo{person}{David Brumley}.}
  \bibinfo{year}{2015}\natexlab{}.
\newblock \showarticletitle{Program-adaptive mutational fuzzing}. In
  \bibinfo{booktitle}{\emph{2015 IEEE Symposium on Security and Privacy}}.
  IEEE, \bibinfo{pages}{725--741}.
\newblock
\urldef\tempurl%
\url{https://doi.org/10.1109/SP.2015.50}
\showDOI{\tempurl}


\bibitem[Chen and Dolan-Gavitt(2025)]%
        {chen2025elfuzz}
\bibfield{author}{\bibinfo{person}{Chuyang Chen} {and} \bibinfo{person}{Brendan
  Dolan-Gavitt}.} \bibinfo{year}{2025}\natexlab{}.
\newblock \showarticletitle{{ELFuzz}: Efficient Input Generation via
  {LLM-driven} Synthesis Over Fuzzer Space}. In \bibinfo{booktitle}{\emph{34th
  USENIX Security Symposium (USENIX Security 25)}}.
  \bibinfo{pages}{6279--6298}.
\newblock


\bibitem[Chen and Chen(2018)]%
        {Chen18-angora}
\bibfield{author}{\bibinfo{person}{Peng Chen} {and} \bibinfo{person}{Hao
  Chen}.} \bibinfo{year}{2018}\natexlab{}.
\newblock \showarticletitle{Angora: Efficient fuzzing by principled search}. In
  \bibinfo{booktitle}{\emph{2018 IEEE Symposium on Security and Privacy (SP)}}.
  IEEE, \bibinfo{pages}{711--725}.
\newblock
\urldef\tempurl%
\url{https://doi.org/10.1109/SP.2018.00046}
\showDOI{\tempurl}


\bibitem[Claessen and Hughes(2000)]%
        {claessen2000quickcheck}
\bibfield{author}{\bibinfo{person}{Koen Claessen} {and} \bibinfo{person}{John
  Hughes}.} \bibinfo{year}{2000}\natexlab{}.
\newblock \showarticletitle{QuickCheck: a lightweight tool for random testing
  of Haskell programs}. In \bibinfo{booktitle}{\emph{Proceedings of the fifth
  ACM SIGPLAN international conference on Functional programming}}.
  \bibinfo{pages}{268--279}.
\newblock


\bibitem[Deng et~al\mbox{.}(2023)]%
        {deng2023large}
\bibfield{author}{\bibinfo{person}{Yinlin Deng},
  \bibinfo{person}{Chunqiu~Steven Xia}, \bibinfo{person}{Chenyuan Yang},
  \bibinfo{person}{Shizhuo~Dylan Zhang}, \bibinfo{person}{Shujing Yang}, {and}
  \bibinfo{person}{Lingming Zhang}.} \bibinfo{year}{2023}\natexlab{}.
\newblock \showarticletitle{Large language models are edge-case fuzzers:
  Testing deep learning libraries via fuzzgpt}.
\newblock \bibinfo{journal}{\emph{arXiv preprint arXiv:2304.02014}}
  (\bibinfo{year}{2023}).
\newblock


\bibitem[Ding and Le~Goues(2021)]%
        {Ding21}
\bibfield{author}{\bibinfo{person}{Zhen~Yu Ding} {and} \bibinfo{person}{Claire
  Le~Goues}.} \bibinfo{year}{2021}\natexlab{}.
\newblock \showarticletitle{An Empirical Study of OSS-Fuzz Bugs}. In
  \bibinfo{booktitle}{\emph{2021 IEEE/ACM 18th International Conference on
  Mining Software Repositories (MSR)}}. IEEE, \bibinfo{publisher}{IEEE Computer
  Society}, \bibinfo{address}{Los Alamitos, CA, USA},
  \bibinfo{pages}{131--142}.
\newblock
\urldef\tempurl%
\url{https://doi.org/10.1109/MSR52588.2021.00026}
\showDOI{\tempurl}


\bibitem[Gan et~al\mbox{.}(2020)]%
        {Gan20}
\bibfield{author}{\bibinfo{person}{Shuitao Gan}, \bibinfo{person}{Chao Zhang},
  \bibinfo{person}{Peng Chen}, \bibinfo{person}{Bodong Zhao},
  \bibinfo{person}{Xiaojun Qin}, \bibinfo{person}{Dong Wu}, {and}
  \bibinfo{person}{Zuoning Chen}.} \bibinfo{year}{2020}\natexlab{}.
\newblock \showarticletitle{{GREYONE}: Data flow sensitive fuzzing}. In
  \bibinfo{booktitle}{\emph{29th USENIX Security Symposium (USENIX Security
  20)}}. \bibinfo{pages}{2577--2594}.
\newblock
\urldef\tempurl%
\url{https://doi.org/10.5555/3489212.3489357}
\showDOI{\tempurl}


\bibitem[Green et~al\mbox{.}(2026)]%
        {green2026stitch}
\bibfield{author}{\bibinfo{person}{Harrison Green}, \bibinfo{person}{Fraser
  Brown}, {and} \bibinfo{person}{Claire Le~Goues}.}
  \bibinfo{year}{2026}\natexlab{}.
\newblock \showarticletitle{Automatic, Expressive, and Scalable Fuzzing with
  Stitching}.
\newblock \bibinfo{journal}{\emph{arXiv preprint arXiv:2602.18689}}
  (\bibinfo{year}{2026}).
\newblock


\bibitem[Guo et~al\mbox{.}(2025)]%
        {guo2025comprehensivesurveybenchmarkssolutions}
\bibfield{author}{\bibinfo{person}{Jiale Guo}, \bibinfo{person}{Suizhi Huang},
  \bibinfo{person}{Mei Li}, \bibinfo{person}{Dong Huang},
  \bibinfo{person}{Xingsheng Chen}, \bibinfo{person}{Regina Zhang},
  \bibinfo{person}{Zhijiang Guo}, \bibinfo{person}{Han Yu},
  \bibinfo{person}{Siu-Ming Yiu}, \bibinfo{person}{Pietro Lio}, {and}
  \bibinfo{person}{Kwok-Yan Lam}.} \bibinfo{year}{2025}\natexlab{}.
\newblock \bibinfo{title}{A Comprehensive Survey on Benchmarks and Solutions in
  Software Engineering of LLM-Empowered Agentic System}.
\newblock
\newblock
\showeprint[arxiv]{2510.09721}~[cs.SE]
\urldef\tempurl%
\url{https://arxiv.org/abs/2510.09721}
\showURL{%
\tempurl}


\bibitem[Holser({[n.\,d.]})]%
        {junit-quickcheck}
\bibfield{author}{\bibinfo{person}{Paul Holser}.}
  \bibinfo{year}{[n.\,d.]}\natexlab{}.
\newblock \bibinfo{title}{junit-quickcheck: Property-based testing,
  JUnit-style}.
\newblock
  \bibinfo{howpublished}{\url{https://github.com/pholser/junit-quickcheck}}.
\newblock
\newblock
\shownote{Accessed: 2021-08-31}.


\bibitem[Hou et~al\mbox{.}(2023)]%
        {hou2023large}
\bibfield{author}{\bibinfo{person}{Xinyi Hou}, \bibinfo{person}{Yanjie Zhao},
  \bibinfo{person}{Yue Liu}, \bibinfo{person}{Zhou Yang},
  \bibinfo{person}{Kailong Wang}, \bibinfo{person}{Li Li},
  \bibinfo{person}{Xiapu Luo}, \bibinfo{person}{David Lo},
  \bibinfo{person}{John Grundy}, {and} \bibinfo{person}{Haoyu Wang}.}
  \bibinfo{year}{2023}\natexlab{}.
\newblock \showarticletitle{Large language models for software engineering: A
  systematic literature review}.
\newblock \bibinfo{journal}{\emph{ACM Transactions on Software Engineering and
  Methodology}} (\bibinfo{year}{2023}).
\newblock


\bibitem[Hough and Bell(2024)]%
        {zeugma}
\bibfield{author}{\bibinfo{person}{Katherine Hough} {and}
  \bibinfo{person}{Jonathan Bell}.} \bibinfo{year}{2024}\natexlab{}.
\newblock \showarticletitle{Crossover in Parametric Fuzzing}. In
  \bibinfo{booktitle}{\emph{Proceedings of the IEEE/ACM 46th International
  Conference on Software Engineering}}. \bibinfo{pages}{1--12}.
\newblock


\bibitem[Ispoglou et~al\mbox{.}(2020)]%
        {ispoglou2020fuzzgen}
\bibfield{author}{\bibinfo{person}{Kyriakos~K Ispoglou},
  \bibinfo{person}{Daniel Austin}, \bibinfo{person}{Vishwath Mohan}, {and}
  \bibinfo{person}{Mathias Payer}.} \bibinfo{year}{2020}\natexlab{}.
\newblock \showarticletitle{Fuzzgen: Automatic fuzzer generation}. In
  \bibinfo{booktitle}{\emph{Proceedings of the 29th USENIX Conference on
  Security Symposium}}. \bibinfo{pages}{2271--2287}.
\newblock


\bibitem[Jeong et~al\mbox{.}(2023)]%
        {jeong2023utopia}
\bibfield{author}{\bibinfo{person}{B. Jeong}, \bibinfo{person}{J. Jang},
  \bibinfo{person}{H. Yi}, \bibinfo{person}{J. Moon}, \bibinfo{person}{J. Kim},
  \bibinfo{person}{I. Jeon}, \bibinfo{person}{T. Kim}, \bibinfo{person}{W.
  Shim}, {and} \bibinfo{person}{Y. Hwang}.} \bibinfo{year}{2023}\natexlab{}.
\newblock \showarticletitle{UTOPIA: Automatic Generation of Fuzz Driver using
  Unit Tests}. In \bibinfo{booktitle}{\emph{2023 2023 IEEE Symposium on
  Security and Privacy (SP) (SP)}}. \bibinfo{publisher}{IEEE Computer Society},
  \bibinfo{address}{Los Alamitos, CA, USA}, \bibinfo{pages}{746--762}.
\newblock
\urldef\tempurl%
\url{https://doi.org/10.1109/SP46215.2023.00043}
\showDOI{\tempurl}


\bibitem[Jiang et~al\mbox{.}(2024)]%
        {jiang2024when}
\bibfield{author}{\bibinfo{person}{Yu Jiang}, \bibinfo{person}{Jie Liang},
  \bibinfo{person}{Fuchen Ma}, \bibinfo{person}{Yuanliang Chen},
  \bibinfo{person}{Chijin Zhou}, \bibinfo{person}{Yuheng Shen},
  \bibinfo{person}{Zhiyong Wu}, \bibinfo{person}{Jingzhou Fu},
  \bibinfo{person}{Mingzhe Wang}, \bibinfo{person}{Shanshan Li}, {and}
  \bibinfo{person}{Quan Zhang}.} \bibinfo{year}{2024}\natexlab{}.
\newblock \showarticletitle{When Fuzzing Meets LLMs: Challenges and
  Opportunities}. In \bibinfo{booktitle}{\emph{Companion Proceedings of the
  32nd ACM International Conference on the Foundations of Software
  Engineering}} (Porto de Galinhas, Brazil) \emph{(\bibinfo{series}{FSE
  2024})}. \bibinfo{publisher}{Association for Computing Machinery},
  \bibinfo{address}{New York, NY, USA}, \bibinfo{pages}{492–496}.
\newblock
\showISBNx{9798400706585}
\urldef\tempurl%
\url{https://doi.org/10.1145/3663529.3663784}
\showDOI{\tempurl}


\bibitem[Jimenez et~al\mbox{.}(2023)]%
        {jimenez2023swe}
\bibfield{author}{\bibinfo{person}{Carlos~E Jimenez}, \bibinfo{person}{John
  Yang}, \bibinfo{person}{Alexander Wettig}, \bibinfo{person}{Shunyu Yao},
  \bibinfo{person}{Kexin Pei}, \bibinfo{person}{Ofir Press}, {and}
  \bibinfo{person}{Karthik Narasimhan}.} \bibinfo{year}{2023}\natexlab{}.
\newblock \showarticletitle{Swe-bench: Can language models resolve real-world
  github issues?}
\newblock \bibinfo{journal}{\emph{arXiv preprint arXiv:2310.06770}}
  (\bibinfo{year}{2023}).
\newblock


\bibitem[Klees et~al\mbox{.}(2018)]%
        {Klees18}
\bibfield{author}{\bibinfo{person}{George Klees}, \bibinfo{person}{Andrew
  Ruef}, \bibinfo{person}{Benji Cooper}, \bibinfo{person}{Shiyi Wei}, {and}
  \bibinfo{person}{Michael Hicks}.} \bibinfo{year}{2018}\natexlab{}.
\newblock \showarticletitle{Evaluating fuzz testing}. In
  \bibinfo{booktitle}{\emph{Proceedings of the 2018 ACM SIGSAC Conference on
  Computer and Communications Security}}. \bibinfo{pages}{2123--2138}.
\newblock
\urldef\tempurl%
\url{https://doi.org/10.1145/3243734.3243804}
\showDOI{\tempurl}


\bibitem[Lahiri et~al\mbox{.}(2022)]%
        {lahiri2022interactive}
\bibfield{author}{\bibinfo{person}{Shuvendu Lahiri}, \bibinfo{person}{Aaditya
  Naik}, \bibinfo{person}{Georgios Sakkas}, \bibinfo{person}{Piali Choudhury},
  \bibinfo{person}{Curtis von Veh}, \bibinfo{person}{Madan Musuvathi},
  \bibinfo{person}{Jeevana~Priya Inala}, \bibinfo{person}{Chenglong Wang},
  {and} \bibinfo{person}{Jianfeng Gao}.} \bibinfo{year}{2022}\natexlab{}.
\newblock \bibinfo{title}{Interactive Code Generation via Test-Driven
  User-Intent Formalization}.
\newblock \bibinfo{howpublished}{arXiv}.
\newblock
\urldef\tempurl%
\url{https://www.microsoft.com/en-us/research/publication/interactive-code-generation-via-test-driven-user-intent-formalization/}
\showURL{%
\tempurl}


\bibitem[Lampropoulos et~al\mbox{.}(2019)]%
        {Lampropoulos19}
\bibfield{author}{\bibinfo{person}{Leonidas Lampropoulos},
  \bibinfo{person}{Michael Hicks}, {and} \bibinfo{person}{Benjamin~C Pierce}.}
  \bibinfo{year}{2019}\natexlab{}.
\newblock \showarticletitle{Coverage guided, property based testing}.
\newblock \bibinfo{journal}{\emph{Proceedings of the ACM on Programming
  Languages}} \bibinfo{volume}{3}, \bibinfo{number}{OOPSLA}, Article
  \bibinfo{articleno}{181} (\bibinfo{date}{Oct.} \bibinfo{year}{2019}),
  \bibinfo{numpages}{29}~pages.
\newblock
\urldef\tempurl%
\url{https://doi.org/10.1145/3360607}
\showDOI{\tempurl}


\bibitem[Lemieux et~al\mbox{.}(2023)]%
        {lemieux2023codamosa}
\bibfield{author}{\bibinfo{person}{Caroline Lemieux},
  \bibinfo{person}{Jeevana~Priya Inala}, \bibinfo{person}{Shuvendu~K Lahiri},
  {and} \bibinfo{person}{Siddhartha Sen}.} \bibinfo{year}{2023}\natexlab{}.
\newblock \showarticletitle{CODAMOSA: Escaping Coverage Plateaus in Test
  Generation with Pre-trained Large Language Models}. In
  \bibinfo{booktitle}{\emph{45th International Conference on Software
  Engineering, ser. ICSE}}.
\newblock


\bibitem[Lemieux and Sen(2018)]%
        {Lemieux18}
\bibfield{author}{\bibinfo{person}{Caroline Lemieux} {and}
  \bibinfo{person}{Koushik Sen}.} \bibinfo{year}{2018}\natexlab{}.
\newblock \showarticletitle{Fairfuzz: A targeted mutation strategy for
  increasing greybox fuzz testing coverage}. In
  \bibinfo{booktitle}{\emph{Proceedings of the 33rd ACM/IEEE International
  Conference on Automated Software Engineering}}. \bibinfo{pages}{475--485}.
\newblock
\urldef\tempurl%
\url{https://doi.org/10.1145/3238147.3238176}
\showDOI{\tempurl}


\bibitem[Liu et~al\mbox{.}(2023)]%
        {ossfuzz23}
\bibfield{author}{\bibinfo{person}{Dongge Liu}, \bibinfo{person}{Jonathan
  Metzman}, {and} \bibinfo{person}{Oliver Chang}.}
  \bibinfo{year}{2023}\natexlab{}.
\newblock \bibinfo{title}{{Open Source Insights}}.
\newblock
  \bibinfo{howpublished}{\url{https://security.googleblog.com/2023/08/ai-powered-fuzzing-breaking-bug-hunting.html}}.
\newblock
\newblock
\shownote{Retrieved February 27, 2024}.


\bibitem[Liu et~al\mbox{.}(2025b)]%
        {liu2025largelanguagemodelbasedagents}
\bibfield{author}{\bibinfo{person}{Junwei Liu}, \bibinfo{person}{Kaixin Wang},
  \bibinfo{person}{Yixuan Chen}, \bibinfo{person}{Xin Peng},
  \bibinfo{person}{Zhenpeng Chen}, \bibinfo{person}{Lingming Zhang}, {and}
  \bibinfo{person}{Yiling Lou}.} \bibinfo{year}{2025}\natexlab{b}.
\newblock \bibinfo{title}{Large Language Model-Based Agents for Software
  Engineering: A Survey}.
\newblock
\newblock
\showeprint[arxiv]{2409.02977}~[cs.SE]
\urldef\tempurl%
\url{https://arxiv.org/abs/2409.02977}
\showURL{%
\tempurl}


\bibitem[Liu et~al\mbox{.}(2024)]%
        {liu2024your}
\bibfield{author}{\bibinfo{person}{Jiawei Liu}, \bibinfo{person}{Chunqiu~Steven
  Xia}, \bibinfo{person}{Yuyao Wang}, {and} \bibinfo{person}{Lingming Zhang}.}
  \bibinfo{year}{2024}\natexlab{}.
\newblock \showarticletitle{Is your code generated by chatgpt really correct?
  rigorous evaluation of large language models for code generation}.
\newblock \bibinfo{journal}{\emph{Advances in Neural Information Processing
  Systems}}  \bibinfo{volume}{36} (\bibinfo{year}{2024}).
\newblock


\bibitem[Liu et~al\mbox{.}(2025a)]%
        {liu2025promefuzz}
\bibfield{author}{\bibinfo{person}{Yuwei Liu}, \bibinfo{person}{Junquan Deng},
  \bibinfo{person}{Xiangkun Jia}, \bibinfo{person}{Yanhao Wang},
  \bibinfo{person}{Minghua Wang}, \bibinfo{person}{Lin Huang},
  \bibinfo{person}{Tao Wei}, {and} \bibinfo{person}{Purui Su}.}
  \bibinfo{year}{2025}\natexlab{a}.
\newblock \showarticletitle{PromeFuzz: A Knowledge-Driven Approach to Fuzzing
  Harness Generation with Large Language Models}. In
  \bibinfo{booktitle}{\emph{Proceedings of the 2025 ACM SIGSAC Conference on
  Computer and Communications Security}}. \bibinfo{pages}{1559--1573}.
\newblock


\bibitem[Lyu et~al\mbox{.}(2019)]%
        {Lyu19}
\bibfield{author}{\bibinfo{person}{Chenyang Lyu}, \bibinfo{person}{Shouling
  Ji}, \bibinfo{person}{Chao Zhang}, \bibinfo{person}{Yuwei Li},
  \bibinfo{person}{Wei-Han Lee}, \bibinfo{person}{Yu Song}, {and}
  \bibinfo{person}{Raheem Beyah}.} \bibinfo{year}{2019}\natexlab{}.
\newblock \showarticletitle{{MOPT}: Optimized mutation scheduling for fuzzers}.
  In \bibinfo{booktitle}{\emph{28th USENIX Security Symposium (USENIX Security
  19)}}. \bibinfo{pages}{1949--1966}.
\newblock
\urldef\tempurl%
\url{https://doi.org/10.5555/3361338.3361473}
\showDOI{\tempurl}


\bibitem[Lyu et~al\mbox{.}(2023)]%
        {lyu2023prompt}
\bibfield{author}{\bibinfo{person}{Yunlong Lyu}, \bibinfo{person}{Yuxuan Xie},
  \bibinfo{person}{Peng Chen}, {and} \bibinfo{person}{Hao Chen}.}
  \bibinfo{year}{2023}\natexlab{}.
\newblock \showarticletitle{Prompt Fuzzing for Fuzz Driver Generation}.
\newblock \bibinfo{journal}{\emph{arXiv preprint arXiv:2312.17677}}
  (\bibinfo{year}{2023}).
\newblock


\bibitem[Maaz et~al\mbox{.}(2025)]%
        {maaz2025agentic}
\bibfield{author}{\bibinfo{person}{Muhammad Maaz}, \bibinfo{person}{Liam
  DeVoe}, \bibinfo{person}{Zac Hatfield-Dodds}, {and} \bibinfo{person}{Nicholas
  Carlini}.} \bibinfo{year}{2025}\natexlab{}.
\newblock \showarticletitle{Agentic Property-Based Testing: Finding Bugs Across
  the Python Ecosystem}.
\newblock \bibinfo{journal}{\emph{arXiv preprint arXiv:2510.09907}}
  (\bibinfo{year}{2025}).
\newblock


\bibitem[Man{\`e}s et~al\mbox{.}(2019)]%
        {Valentin19}
\bibfield{author}{\bibinfo{person}{Valentin Jean~Marie Man{\`e}s},
  \bibinfo{person}{HyungSeok Han}, \bibinfo{person}{Choongwoo Han},
  \bibinfo{person}{Sang~Kil Cha}, \bibinfo{person}{Manuel Egele},
  \bibinfo{person}{Edward~J Schwartz}, {and} \bibinfo{person}{Maverick Woo}.}
  \bibinfo{year}{2019}\natexlab{}.
\newblock \showarticletitle{The art, science, and engineering of fuzzing: A
  survey}.
\newblock \bibinfo{journal}{\emph{IEEE Transactions on Software Engineering}}
  (\bibinfo{year}{2019}).
\newblock
\urldef\tempurl%
\url{https://doi.org/10.1109/TSE.2019.2946563}
\showDOI{\tempurl}


\bibitem[Meng et~al\mbox{.}(2024)]%
        {meng2024large}
\bibfield{author}{\bibinfo{person}{Ruijie Meng}, \bibinfo{person}{Martin
  Mirchev}, \bibinfo{person}{Marcel B{\"o}hme}, {and} \bibinfo{person}{Abhik
  Roychoudhury}.} \bibinfo{year}{2024}\natexlab{}.
\newblock \showarticletitle{Large language model guided protocol fuzzing}. In
  \bibinfo{booktitle}{\emph{Proceedings of the 31st Annual Network and
  Distributed System Security Symposium (NDSS)}}.
\newblock


\bibitem[Miller et~al\mbox{.}(1990)]%
        {Miller90}
\bibfield{author}{\bibinfo{person}{Barton~P. Miller}, \bibinfo{person}{Louis
  Fredriksen}, {and} \bibinfo{person}{Bryan So}.}
  \bibinfo{year}{1990}\natexlab{}.
\newblock \showarticletitle{An Empirical Study of the Reliability of UNIX
  Utilities}.
\newblock \bibinfo{journal}{\emph{Commun. ACM}} \bibinfo{volume}{33},
  \bibinfo{number}{12} (\bibinfo{date}{dec} \bibinfo{year}{1990}),
  \bibinfo{pages}{32–44}.
\newblock
\showISSN{0001-0782}
\urldef\tempurl%
\url{https://doi.org/10.1145/96267.96279}
\showDOI{\tempurl}


\bibitem[Nguyen and Grunske(2022)]%
        {nguyen2022bedivfuzz}
\bibfield{author}{\bibinfo{person}{Hoang~Lam Nguyen} {and}
  \bibinfo{person}{Lars Grunske}.} \bibinfo{year}{2022}\natexlab{}.
\newblock \showarticletitle{Bedivfuzz: Integrating behavioral diversity into
  generator-based fuzzing}. In \bibinfo{booktitle}{\emph{Proceedings of the
  44th International Conference on Software Engineering}}.
  \bibinfo{pages}{249--261}.
\newblock


\bibitem[Padhye et~al\mbox{.}(2019a)]%
        {Padhye19-jqf}
\bibfield{author}{\bibinfo{person}{Rohan Padhye}, \bibinfo{person}{Caroline
  Lemieux}, {and} \bibinfo{person}{Koushik Sen}.}
  \bibinfo{year}{2019}\natexlab{a}.
\newblock \showarticletitle{{JQF}: Coverage-guided Property-based Testing in
  {Java}}. In \bibinfo{booktitle}{\emph{Proceedings of the 28th ACM SIGSOFT
  International Symposium on Software Testing and Analysis}} (Beijing, China)
  \emph{(\bibinfo{series}{ISSTA'19})}. \bibinfo{publisher}{Association for
  Computing Machinery}, \bibinfo{address}{New York, NY, USA},
  \bibinfo{pages}{398--401}.
\newblock
\showISBNx{978-1-4503-6224-5}
\urldef\tempurl%
\url{https://doi.org/10.1145/3293882.3339002}
\showDOI{\tempurl}


\bibitem[Padhye et~al\mbox{.}(2019b)]%
        {Padhye19-zest}
\bibfield{author}{\bibinfo{person}{Rohan Padhye}, \bibinfo{person}{Caroline
  Lemieux}, \bibinfo{person}{Koushik Sen}, \bibinfo{person}{Mike Papadakis},
  {and} \bibinfo{person}{Yves Le~Traon}.} \bibinfo{year}{2019}\natexlab{b}.
\newblock \showarticletitle{Semantic Fuzzing with {Zest}}. In
  \bibinfo{booktitle}{\emph{Proceedings of the 28th ACM SIGSOFT International
  Symposium on Software Testing and Analysis}} (Beijing, China)
  \emph{(\bibinfo{series}{ISSTA 2019})}. \bibinfo{publisher}{ACM},
  \bibinfo{pages}{329--340}.
\newblock
\showISBNx{978-1-4503-6224-5}
\urldef\tempurl%
\url{https://doi.org/10.1145/3293882.3330576}
\showDOI{\tempurl}


\bibitem[Padhye et~al\mbox{.}(2019c)]%
        {Padhye19-chocopy}
\bibfield{author}{\bibinfo{person}{Rohan Padhye}, \bibinfo{person}{Koushik
  Sen}, {and} \bibinfo{person}{Paul~N. Hilfinger}.}
  \bibinfo{year}{2019}\natexlab{c}.
\newblock \showarticletitle{ChocoPy: A Programming Language for Compilers
  Courses}. In \bibinfo{booktitle}{\emph{Proceedings of the 2019 ACM SIGPLAN
  Symposium on SPLASH-E}} (Athens, Greece) \emph{(\bibinfo{series}{SPLASH-E
  2019})}. \bibinfo{publisher}{Association for Computing Machinery},
  \bibinfo{pages}{41–45}.
\newblock
\showISBNx{9781450369893}
\urldef\tempurl%
\url{https://doi.org/10.1145/3358711.3361627}
\showDOI{\tempurl}


\bibitem[Rao et~al\mbox{.}(2024)]%
        {rao2024diffspec}
\bibfield{author}{\bibinfo{person}{Nikitha Rao}, \bibinfo{person}{Elizabeth
  Gilbert}, \bibinfo{person}{Tahina Ramananandro}, \bibinfo{person}{Nikhil
  Swamy}, \bibinfo{person}{Claire~Le Goues}, {and} \bibinfo{person}{Sarah
  Fakhoury}.} \bibinfo{year}{2024}\natexlab{}.
\newblock \showarticletitle{DiffSpec: Differential Testing with LLMs using
  Natural Language Specifications and Code Artifacts}.
\newblock \bibinfo{journal}{\emph{arXiv preprint arXiv:2410.04249}}
  (\bibinfo{year}{2024}).
\newblock


\bibitem[Rao et~al\mbox{.}(2023)]%
        {rao2023cat}
\bibfield{author}{\bibinfo{person}{Nikitha Rao}, \bibinfo{person}{Kush Jain},
  \bibinfo{person}{Uri Alon}, \bibinfo{person}{Claire Le~Goues}, {and}
  \bibinfo{person}{Vincent~J Hellendoorn}.} \bibinfo{year}{2023}\natexlab{}.
\newblock \showarticletitle{CAT-LM training language models on aligned code and
  tests}. In \bibinfo{booktitle}{\emph{2023 38th IEEE/ACM International
  Conference on Automated Software Engineering (ASE)}}. IEEE,
  \bibinfo{pages}{409--420}.
\newblock


\bibitem[Rebert et~al\mbox{.}(2014)]%
        {Rebert14}
\bibfield{author}{\bibinfo{person}{Alexandre Rebert}, \bibinfo{person}{Sang~Kil
  Cha}, \bibinfo{person}{Thanassis Avgerinos}, \bibinfo{person}{Jonathan
  Foote}, \bibinfo{person}{David Warren}, \bibinfo{person}{Gustavo Grieco},
  {and} \bibinfo{person}{David Brumley}.} \bibinfo{year}{2014}\natexlab{}.
\newblock \showarticletitle{Optimizing seed selection for fuzzing}. In
  \bibinfo{booktitle}{\emph{23rd USENIX Security Symposium (USENIX Security
  14)}}. \bibinfo{pages}{861--875}.
\newblock
\urldef\tempurl%
\url{https://doi.org/10.5555/2671225.2671280}
\showDOI{\tempurl}


\bibitem[Regehr et~al\mbox{.}(2012)]%
        {Regehr12}
\bibfield{author}{\bibinfo{person}{John Regehr}, \bibinfo{person}{Yang Chen},
  \bibinfo{person}{Pascal Cuoq}, \bibinfo{person}{Eric Eide},
  \bibinfo{person}{Chucky Ellison}, {and} \bibinfo{person}{Xuejun Yang}.}
  \bibinfo{year}{2012}\natexlab{}.
\newblock \showarticletitle{Test-Case Reduction for {C} Compiler Bugs}. In
  \bibinfo{booktitle}{\emph{Proceedings of the 33rd ACM SIGPLAN Conference on
  Programming Language Design and Implementation}} (Beijing, China)
  \emph{(\bibinfo{series}{PLDI ’12})}. \bibinfo{publisher}{Association for
  Computing Machinery}, \bibinfo{address}{New York, NY, USA},
  \bibinfo{pages}{335–346}.
\newblock
\showISBNx{9781450312059}
\urldef\tempurl%
\url{https://doi.org/10.1145/2254064.2254104}
\showDOI{\tempurl}


\bibitem[Sch{\"a}fer et~al\mbox{.}(2023)]%
        {schafer2023empirical}
\bibfield{author}{\bibinfo{person}{Max Sch{\"a}fer}, \bibinfo{person}{Sarah
  Nadi}, \bibinfo{person}{Aryaz Eghbali}, {and} \bibinfo{person}{Frank Tip}.}
  \bibinfo{year}{2023}\natexlab{}.
\newblock \showarticletitle{An empirical evaluation of using large language
  models for automated unit test generation}.
\newblock \bibinfo{journal}{\emph{IEEE Transactions on Software Engineering}}
  (\bibinfo{year}{2023}).
\newblock


\bibitem[Schäfer et~al\mbox{.}(2023)]%
        {schafer2023testpilot}
\bibfield{author}{\bibinfo{person}{Max Schäfer}, \bibinfo{person}{Sarah Nadi},
  \bibinfo{person}{Aryaz Eghbali}, {and} \bibinfo{person}{Frank Tip}.}
  \bibinfo{year}{2023}\natexlab{}.
\newblock \bibinfo{title}{Adaptive Test Generation Using a Large Language
  Model}.
\newblock
\newblock
\showeprint[arxiv]{2302.06527}~[cs.SE]


\bibitem[Vikram et~al\mbox{.}(2023a)]%
        {mu2}
\bibfield{author}{\bibinfo{person}{Vasudev Vikram}, \bibinfo{person}{Isabella
  Laybourn}, \bibinfo{person}{Ao Li}, \bibinfo{person}{Nicole Nair},
  \bibinfo{person}{Kelton OBrien}, \bibinfo{person}{Rafaello Sanna}, {and}
  \bibinfo{person}{Rohan Padhye}.} \bibinfo{year}{2023}\natexlab{a}.
\newblock \showarticletitle{Guiding Greybox Fuzzing with Mutation Testing}. In
  \bibinfo{booktitle}{\emph{Proceedings of the 32nd ACM SIGSOFT International
  Symposium on Software Testing and Analysis}} (, Seattle, WA, USA,)
  \emph{(\bibinfo{series}{ISSTA 2023})}. \bibinfo{publisher}{Association for
  Computing Machinery}, \bibinfo{address}{New York, NY, USA},
  \bibinfo{pages}{929–941}.
\newblock
\showISBNx{9798400702211}
\urldef\tempurl%
\url{https://doi.org/10.1145/3597926.3598107}
\showDOI{\tempurl}


\bibitem[Vikram et~al\mbox{.}(2023b)]%
        {vikram2023can}
\bibfield{author}{\bibinfo{person}{Vasudev Vikram}, \bibinfo{person}{Caroline
  Lemieux}, \bibinfo{person}{Joshua Sunshine}, {and} \bibinfo{person}{Rohan
  Padhye}.} \bibinfo{year}{2023}\natexlab{b}.
\newblock \showarticletitle{Can large language models write good property-based
  tests?}
\newblock \bibinfo{journal}{\emph{arXiv preprint arXiv:2307.04346}}
  (\bibinfo{year}{2023}).
\newblock


\bibitem[Vikram et~al\mbox{.}(2021)]%
        {Vikram21}
\bibfield{author}{\bibinfo{person}{Vasudev Vikram}, \bibinfo{person}{Rohan
  Padhye}, {and} \bibinfo{person}{Koushik Sen}.}
  \bibinfo{year}{2021}\natexlab{}.
\newblock \showarticletitle{Growing A Test Corpus with Bonsai Fuzzing}. In
  \bibinfo{booktitle}{\emph{43rd {IEEE/ACM} International Conference on
  Software Engineering, {ICSE} 2021, Madrid, Spain, 22-30 May 2021}}.
  \bibinfo{publisher}{{IEEE}}, \bibinfo{pages}{723--735}.
\newblock
\urldef\tempurl%
\url{https://doi.org/10.1109/ICSE43902.2021.00072}
\showDOI{\tempurl}


\bibitem[Wang et~al\mbox{.}(2024)]%
        {wang2024openhands}
\bibfield{author}{\bibinfo{person}{Xingyao Wang}, \bibinfo{person}{Boxuan Li},
  \bibinfo{person}{Yufan Song}, \bibinfo{person}{Frank~F Xu},
  \bibinfo{person}{Xiangru Tang}, \bibinfo{person}{Mingchen Zhuge},
  \bibinfo{person}{Jiayi Pan}, \bibinfo{person}{Yueqi Song},
  \bibinfo{person}{Bowen Li}, \bibinfo{person}{Jaskirat Singh},
  {et~al\mbox{.}}} \bibinfo{year}{2024}\natexlab{}.
\newblock \showarticletitle{OpenHands: An Open Platform for AI Software
  Developers as Generalist Agents}.
\newblock \bibinfo{journal}{\emph{arXiv preprint arXiv:2407.16741}}
  (\bibinfo{year}{2024}).
\newblock


\bibitem[Wu et~al\mbox{.}(2022)]%
        {havoc}
\bibfield{author}{\bibinfo{person}{Mingyuan Wu}, \bibinfo{person}{Ling Jiang},
  \bibinfo{person}{Jiahong Xiang}, \bibinfo{person}{Yanwei Huang},
  \bibinfo{person}{Heming Cui}, \bibinfo{person}{Lingming Zhang}, {and}
  \bibinfo{person}{Yuqun Zhang}.} \bibinfo{year}{2022}\natexlab{}.
\newblock \showarticletitle{One Fuzzing Strategy to Rule Them All}. In
  \bibinfo{booktitle}{\emph{Proceedings of the 44th International Conference on
  Software Engineering}} (Pittsburgh, Pennsylvania)
  \emph{(\bibinfo{series}{ICSE '22})}. \bibinfo{publisher}{Association for
  Computing Machinery}, \bibinfo{address}{New York, NY, USA},
  \bibinfo{pages}{1634–1645}.
\newblock
\showISBNx{9781450392211}
\urldef\tempurl%
\url{https://doi.org/10.1145/3510003.3510174}
\showDOI{\tempurl}


\bibitem[Xia et~al\mbox{.}(2024)]%
        {xia2024fuzz4all}
\bibfield{author}{\bibinfo{person}{Chunqiu~Steven Xia}, \bibinfo{person}{Matteo
  Paltenghi}, \bibinfo{person}{Jia Le~Tian}, \bibinfo{person}{Michael Pradel},
  {and} \bibinfo{person}{Lingming Zhang}.} \bibinfo{year}{2024}\natexlab{}.
\newblock \showarticletitle{Fuzz4all: Universal fuzzing with large language
  models}.
\newblock \bibinfo{journal}{\emph{Proc. IEEE/ACM ICSE}} (\bibinfo{year}{2024}).
\newblock


\bibitem[Xia et~al\mbox{.}(2025)]%
        {xia2025livesweagentsoftwareengineeringagents}
\bibfield{author}{\bibinfo{person}{Chunqiu~Steven Xia}, \bibinfo{person}{Zhe
  Wang}, \bibinfo{person}{Yan Yang}, \bibinfo{person}{Yuxiang Wei}, {and}
  \bibinfo{person}{Lingming Zhang}.} \bibinfo{year}{2025}\natexlab{}.
\newblock \bibinfo{title}{Live-SWE-agent: Can Software Engineering Agents
  Self-Evolve on the Fly?}
\newblock
\newblock
\showeprint[arxiv]{2511.13646}~[cs.SE]
\urldef\tempurl%
\url{https://arxiv.org/abs/2511.13646}
\showURL{%
\tempurl}


\bibitem[Xu et~al\mbox{.}(2024)]%
        {xu2024ckgfuzzer}
\bibfield{author}{\bibinfo{person}{Hanxiang Xu}, \bibinfo{person}{Wei Ma},
  \bibinfo{person}{Ting Zhou}, \bibinfo{person}{Yanjie Zhao},
  \bibinfo{person}{Kai Chen}, \bibinfo{person}{Qiang Hu}, \bibinfo{person}{Yang
  Liu}, {and} \bibinfo{person}{Haoyu Wang}.} \bibinfo{year}{2024}\natexlab{}.
\newblock \showarticletitle{{CKGFuzzer}: {LLM}-Based Fuzz Driver Generation
  Enhanced By Code Knowledge Graph}.
\newblock \bibinfo{journal}{\emph{arXiv preprint arXiv:2411.11532}}
  (\bibinfo{year}{2024}).
\newblock


\bibitem[Yang et~al\mbox{.}(2023)]%
        {yang2023whitefox}
\bibfield{author}{\bibinfo{person}{Chenyuan Yang}, \bibinfo{person}{Yinlin
  Deng}, \bibinfo{person}{Runyu Lu}, \bibinfo{person}{Jiayi Yao},
  \bibinfo{person}{Jiawei Liu}, \bibinfo{person}{Reyhaneh Jabbarvand}, {and}
  \bibinfo{person}{Lingming Zhang}.} \bibinfo{year}{2023}\natexlab{}.
\newblock \showarticletitle{WhiteFox: White-Box Compiler Fuzzing Empowered by
  Large Language Models}.
\newblock \bibinfo{journal}{\emph{arXiv preprint arXiv:2310.15991}}
  (\bibinfo{year}{2023}).
\newblock


\bibitem[Yang et~al\mbox{.}(2024)]%
        {yang2024swe}
\bibfield{author}{\bibinfo{person}{John Yang}, \bibinfo{person}{Carlos~E
  Jimenez}, \bibinfo{person}{Alexander Wettig}, \bibinfo{person}{Kilian
  Lieret}, \bibinfo{person}{Shunyu Yao}, \bibinfo{person}{Karthik Narasimhan},
  {and} \bibinfo{person}{Ofir Press}.} \bibinfo{year}{2024}\natexlab{}.
\newblock \showarticletitle{Swe-agent: Agent-computer interfaces enable
  automated software engineering}.
\newblock \bibinfo{journal}{\emph{arXiv preprint arXiv:2405.15793}}
  (\bibinfo{year}{2024}).
\newblock


\bibitem[Yang et~al\mbox{.}(2011)]%
        {Yang11}
\bibfield{author}{\bibinfo{person}{Xuejun Yang}, \bibinfo{person}{Yang Chen},
  \bibinfo{person}{Eric Eide}, {and} \bibinfo{person}{John Regehr}.}
  \bibinfo{year}{2011}\natexlab{}.
\newblock \showarticletitle{Finding and Understanding Bugs in {C} Compilers}.
  In \bibinfo{booktitle}{\emph{Proceedings of the 32nd ACM SIGPLAN Conference
  on Programming Language Design and Implementation}}
  \emph{(\bibinfo{series}{PLDI '11})}.
\newblock


\bibitem[Zalewski(2014)]%
        {Afl}
\bibfield{author}{\bibinfo{person}{Michal Zalewski}.}
  \bibinfo{year}{2014}\natexlab{}.
\newblock \bibinfo{title}{American Fuzzy Lop}.
\newblock \bibinfo{howpublished}{\url{https://lcamtuf.coredump.cx/afl/}}.
\newblock
\newblock
\shownote{Accessed February 11, 2022}.


\bibitem[Zhang et~al\mbox{.}(2023)]%
        {zhang2023understanding}
\bibfield{author}{\bibinfo{person}{Cen Zhang}, \bibinfo{person}{Mingqiang Bai},
  \bibinfo{person}{Yaowen Zheng}, \bibinfo{person}{Yeting Li},
  \bibinfo{person}{Xiaofei Xie}, \bibinfo{person}{Yuekang Li},
  \bibinfo{person}{Wei Ma}, \bibinfo{person}{Limin Sun}, {and}
  \bibinfo{person}{Yang Liu}.} \bibinfo{year}{2023}\natexlab{}.
\newblock \showarticletitle{Understanding large language model based fuzz
  driver generation}.
\newblock \bibinfo{journal}{\emph{arXiv preprint arXiv:2307.12469}}
  (\bibinfo{year}{2023}).
\newblock


\bibitem[Zhang et~al\mbox{.}(2025)]%
        {zhang2025low}
\bibfield{author}{\bibinfo{person}{Kunpeng Zhang}, \bibinfo{person}{Zongjie
  Li}, \bibinfo{person}{Daoyuan Wu}, \bibinfo{person}{Shuai Wang}, {and}
  \bibinfo{person}{Xin Xia}.} \bibinfo{year}{2025}\natexlab{}.
\newblock \showarticletitle{{Low-Cost} and Comprehensive Non-textual Input
  Fuzzing with {LLM-Synthesized} Input Generators}. In
  \bibinfo{booktitle}{\emph{34th USENIX Security Symposium (USENIX Security
  25)}}. \bibinfo{pages}{6999--7018}.
\newblock


\end{thebibliography}
